\documentclass{article} 
\usepackage{iclr2025_conference,times}

\usepackage{amsmath,amsfonts,bm}

\def\eqref#1{equation~\ref{#1}}

\def\1{\bm{1}}

\DeclareMathAlphabet{\mathsfit}{\encodingdefault}{\sfdefault}{m}{sl}
\SetMathAlphabet{\mathsfit}{bold}{\encodingdefault}{\sfdefault}{bx}{n}

\usepackage{hyperref}
\usepackage{url}
\usepackage{multirow}
\usepackage{bigfoot}
\usepackage{adjustbox}
\usepackage{booktabs}
\usepackage{graphicx}   
\usepackage{subcaption} 
\usepackage{caption}
\usepackage[colorinlistoftodos, textwidth=3cm]{todonotes}
\usepackage{comment}
\usepackage{float}
\usepackage[table]{xcolor}
\usepackage{tikz}
\usetikzlibrary{positioning,arrows.meta,fit,calc}

\ifdefined\headwidth\fi
\newcommand{\logoorelse}[3]{\IfFileExists{images/#1.pdf}{#2}{\IfFileExists{images/#1.png}{#2}{#3}}}
\newcommand{\ibmlogo}{\logoorelse{ibm-logo}{\includegraphics[height=0.72em]{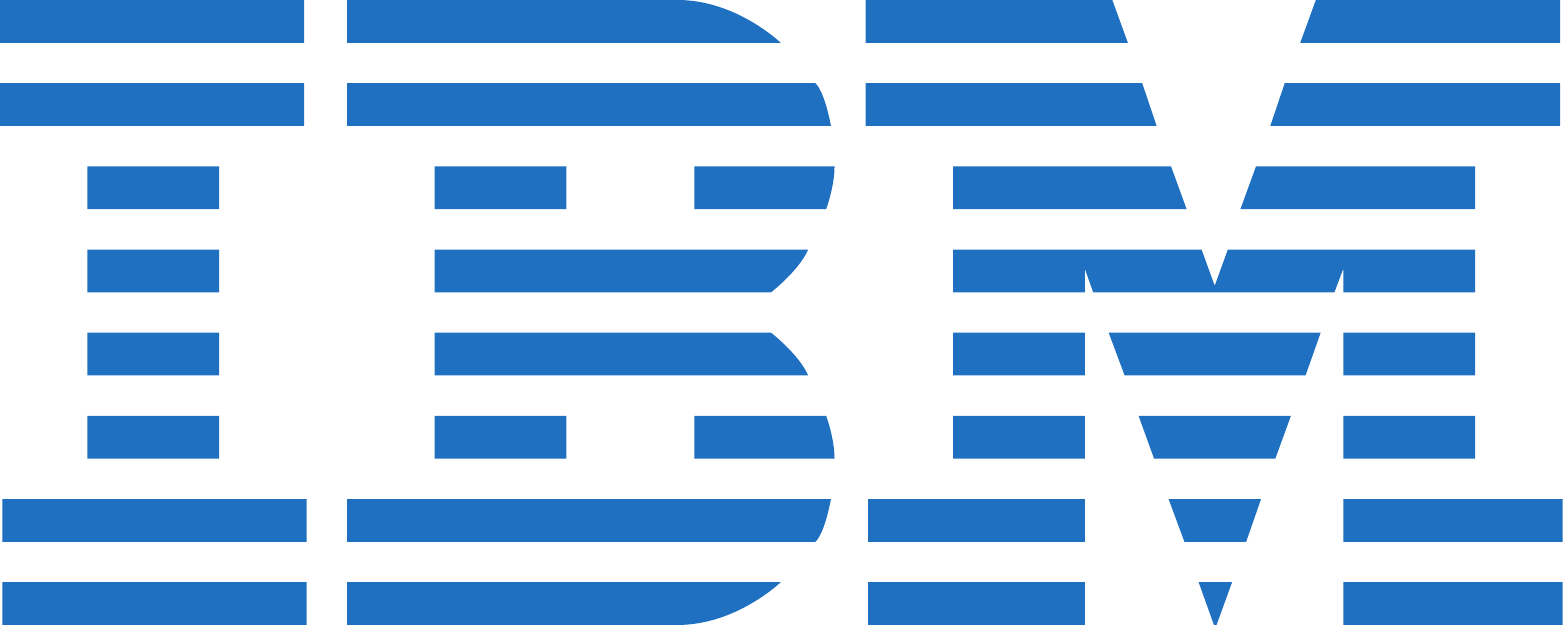}}{\textbf{IBM}}}
\newcommand{\granitelogo}{\logoorelse{granite-embedding-logo-small}{\raisebox{-0.2em}{\includegraphics[height=1.1em]{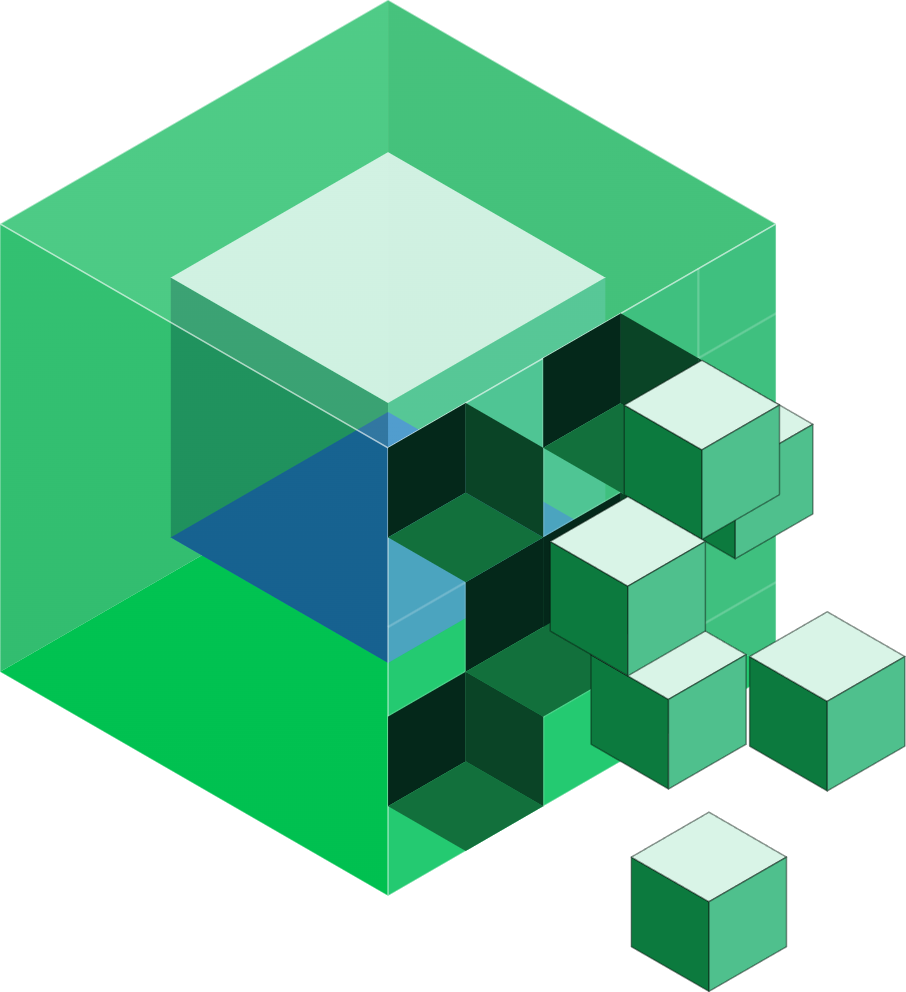}}\hspace{0.4em}}{}}
\newcommand{\ibmresearch}{\ibmlogo\hspace{0.35em}Research}
\newcommand{\graniteembeddings}{\granitelogo Granite Embeddings}

\makeatletter
\renewcommand{\section}{\@startsection{section}{1}{\z@}{-2.0ex plus -0.5ex minus -.2ex}{1.5ex plus 0.3ex minus .2ex}{\large\bfseries\raggedright}}
\renewcommand{\subsection}{\@startsection{subsection}{2}{\z@}{-1.8ex plus -0.5ex minus -.2ex}{0.8ex plus .2ex}{\normalsize\bfseries\raggedright}}
\renewcommand{\subsubsection}{\@startsection{subsubsection}{3}{\z@}{-1.5ex plus -0.5ex minus -.2ex}{0.5ex plus .2ex}{\normalsize\bfseries\raggedright}}
\renewcommand{\@seccntformat}[1]{\csname the#1\endcsname.\quad}
\makeatother

\newcommand{\verify}[2][W?]{\todo[inline,color=orange!30]{\textbf{#1} VERIFY: #2}}
\newcommand{\modelfullname}{\texttt{granite-embedding-748m-vision-r2}}
\newcommand{\modelname}{GraniteEmb-748m}
\newcommand{\textmodelname}{GraniteEmb-Text-311m}

\title{Granite Embedding Vision R2}

\author{Granite Embedding Team,
IBM Research\thanks{See Section \ref{sec:contributions} for full author list. For questions, comments, compliments contact raduf@us.ibm.com.} \thanks{For feedback or comments on this work, please open an issue at \url{https://github.com/ibm-granite/granite-embedding-models}.}
}

\iclrfinalcopy 
\begin{document}

\maketitle
\lhead{\ibmresearch}
\rhead{\graniteembeddings}

\begin{abstract}
We introduce \modelfullname{}, a 748M-parameter multilingual embedding model that makes visual document retrieval practical at enterprise scale. It retrieves page images directly from text queries without OCR or layout analysis, producing both single-vector embeddings for fast search and compact multi-vector embeddings for accurate late interaction in one forward pass. Across eight visual document retrieval benchmarks, it averages 0.721 nDCG while encoding about 35 pages per second on one NVIDIA H200 GPU-faster than all 20 comparison models in our evaluation, placing it on the speed--quality Pareto front. The only evaluated model under 1B parameters with higher retrieval quality runs at half its speed and stores roughly 1{,}500 vectors per page, compared with about 600 for ours. The model also matches the text-only granite-embedding-311m-multilingual-r2 on multilingual text retrieval and can query existing text indexes built with it, allowing visual retrieval to complement established text collections. Trained on data reviewed through IBM's data governance process, it is released under the Apache 2.0 license at \url{https://huggingface.co/collections/ibm-granite}.
\end{abstract}


\begin{figure}[ht]
    \centering
    \begin{subfigure}[t]{0.495\linewidth}
        \centering
        \includegraphics[width=\linewidth]{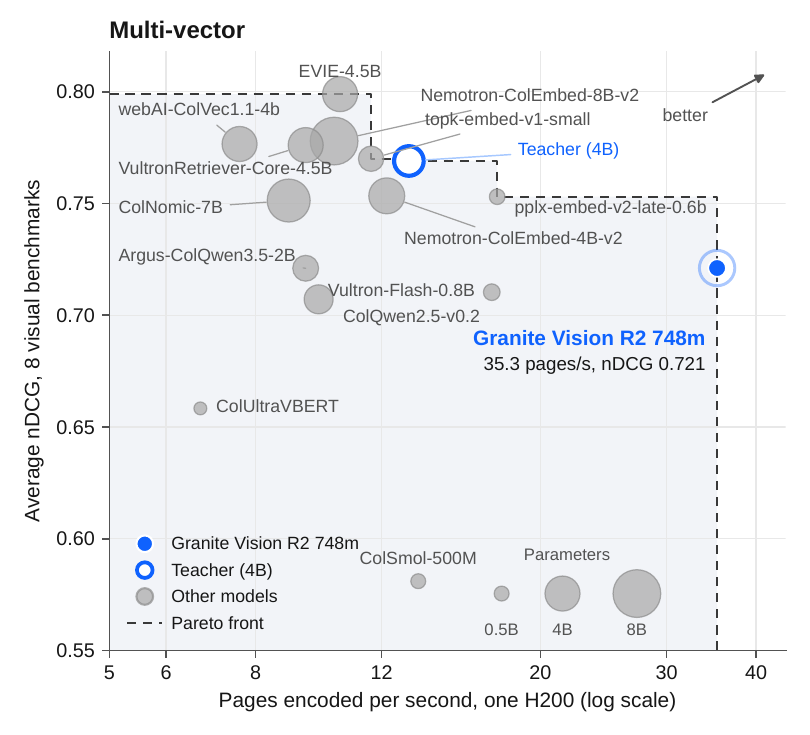}
        \caption{Multi-vector models.}
        \label{fig:pareto-mv}
    \end{subfigure}\hfill
    \begin{subfigure}[t]{0.495\linewidth}
        \centering
        \includegraphics[width=\linewidth]{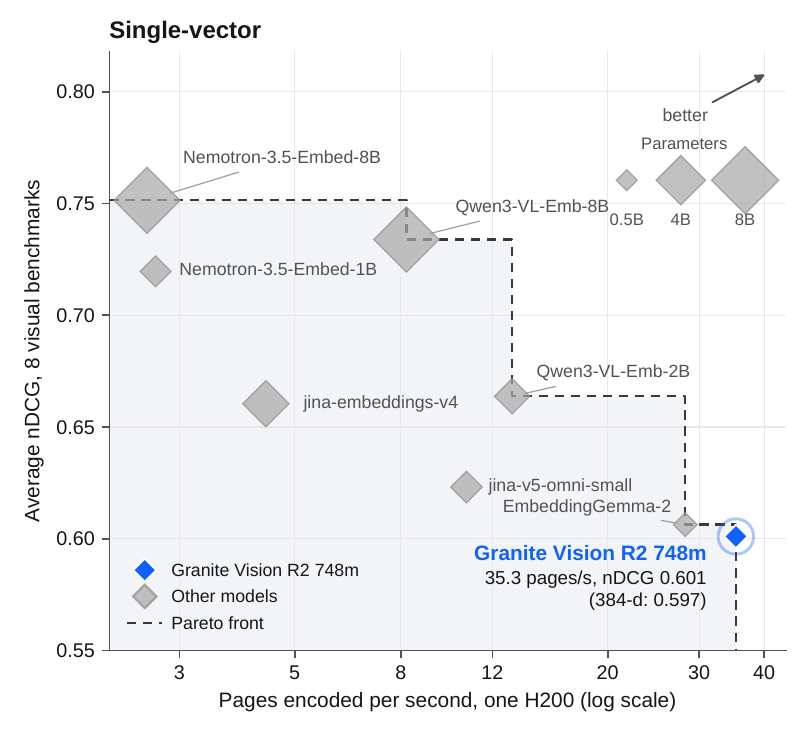}
        \caption{Single-vector models.}
        \label{fig:pareto-sv}
    \end{subfigure}
    \caption{Speed vs.\ accuracy for visual document retrievers. The y-axis is the average nDCG over the eight benchmarks of Table~\ref{tab:visual-results}; the x-axis is page images encoded per second on one H200 (log scale). Marker area is proportional to parameter count. The dashed line is the Pareto front over the released models; the teacher is shown for reference (multi-vector only). Data in Table~\ref{tab:speed}.}
    \label{fig:pareto}
\end{figure}

\section{Introduction}
\label{sec:intro}

Enterprise knowledge lives in documents: reports, slide decks, scanned forms, financial statements and manuals whose meaning is carried as much by tables, charts and layout as by running text. Doing text-only retrieval over such collections requires a preprocessing pipeline of OCR, layout detection, table extraction and chunking that is slow and loses the information stored in visual structure: layout detection and OCR alone take about 3.5\,s per page, ten times as long as encoding the page image with a visual retriever \citep{faysse2025colpali}. We sidestep this pipeline with visual document retrieval: a vision-language model embeds each page image directly, and a text query is matched against those page embeddings \citep{faysse2025colpali}.

Strong visual retrievers are expensive at enterprise scale. Of the 20 open retrievers we compare, 15 have more than 1.5B parameters, and most use late interaction \citep{khattab2020colbert}, storing one vector per image patch or token, so the multi-vector index can be hundreds of times larger than a single-vector one. Encoding speed sets how quickly a collection can be indexed and kept current: at the 2.6 to 17.4 pages per second of the models that score higher than ours, one million pages take 16 to 107 hours on one H200. Enterprise collections are also multilingual, and the smallest retrievers are weak there: the two smallest multi-vector models we compare score below 0.30 nDCG on M3DR, against 0.548 for ours.

This report introduces \modelfullname{} (\modelname{} for short), which builds on the Granite Embedding Multilingual R2 text encoder \citep{awasthy2026graniteembeddingmultilingualr2} to bring visual document retrieval to a compact, multilingual model. Its main properties are:
\begin{itemize}
    \item \textbf{Fast, on the speed--quality Pareto front.} It encodes 35.3 pages/s on one H200 (about 8 hours per million pages), the fastest of the 20 retrievers we compared, and every model that scores higher is slower; the only one under 1B parameters is half as fast (Section~\ref{subsec:speed}).
    \item \textbf{Accurate for its size, with a small index.} Its multi-vector head averages 0.721 nDCG over eight visual document retrieval benchmarks, ahead of the closest model in size (0.710) with 2.5$\times$ narrower token vectors (128 against 320 dimensions), and it keeps 144 to 720 vectors per page against about 1{,}500 for the strongest model under 1B parameters (Section~\ref{subsec:visual-results}).
    \item \textbf{Hybrid search from one model, with or without a multi-vector index.} A single model produces both single-vector and multi-vector representations. The single vector shortlists pages and the model's own multi-vectors rerank them, within 0.001 nDCG@5 of exhaustive multi-vector search for about 1.5\,ms more per query. The multi-vectors can be stored, at about 100$\times$ the size of the single vector, or recomputed for the shortlist only, which keeps just the single vector and the page image and costs 0.2--0.4\,s per query at $k=10$ (Section~\ref{sec:search}).
    \item \textbf{Adjustable index size.} Both outputs shrink at index time without retraining: token pooling keeps 98.8\% of quality at 128 vectors per page, and the single vector truncated to 384 dimensions loses only 0.004 on visual and 0.001 on text retrieval (Section~\ref{sec:search}).
    \item \textbf{Multilingual, and still a text retriever.} Its multi-vector head matches the single-vector, text-only granite-embedding-311m-multilingual-r2 (\textmodelname{} for short) on multilingual text retrieval (0.656 against 0.652), and its queries can search an existing text index of that model as well as its own (Section~\ref{subsec:text-results}, Appendix~\ref{app:text-compat}).
\end{itemize}

\begin{figure}[t]
    \centering
    \includegraphics[width=\textwidth]{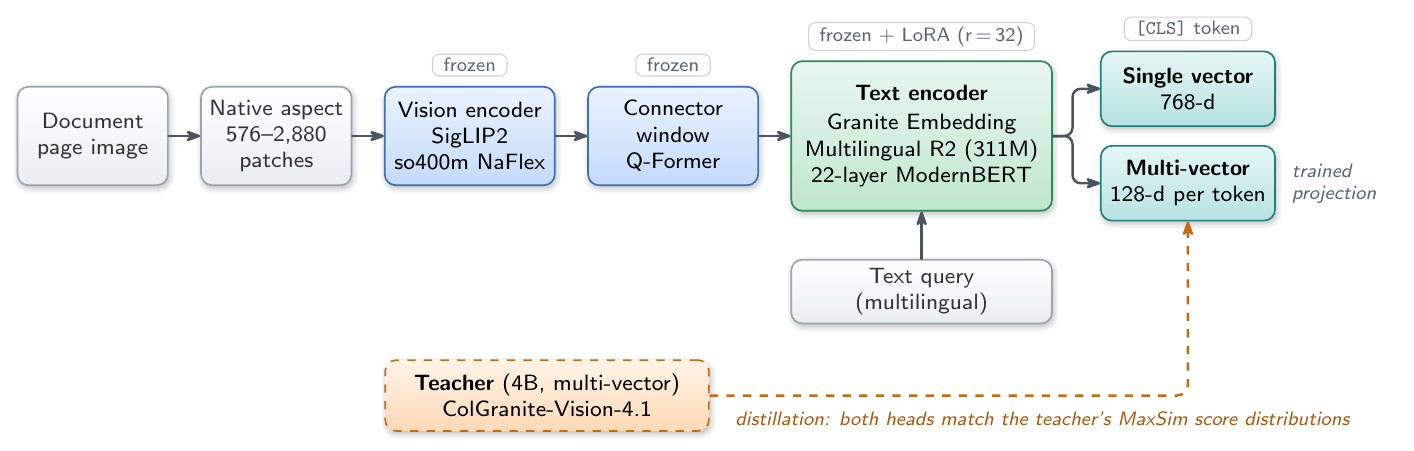}
    \caption{Architecture of \modelname. A NaFlex vision tower encodes page images at close to native aspect ratio; a window Q-Former compresses the patches into the text encoder's input space. The final hidden states feed a single-vector (\texttt{[CLS]}) and a 128-dimensional multi-vector head. Tags mark what is frozen during distillation.}
    \label{fig:architecture}
\end{figure}

\section{Related Work}
\label{sec:related}

\paragraph{Visual document retrieval.} ColPali \citep{faysse2025colpali} showed that a vision-language model trained with a late-interaction objective can retrieve document pages from their images and outperform OCR-based pipelines. Its authors also introduced the ViDoRe benchmark, since extended with harder, multilingual and enterprise-domain tasks \citep{vidorev3}. Subsequent models broadened coverage to multilingual and multi-task settings \citep{gunther2025jinav4,moca2025}. ModernVBERT \citep{modernvbert2025} showed that a compact bidirectional encoder of roughly 250M parameters, aligned to a vision encoder with masked language modeling, can approach much larger decoder-based retrievers; our modality-alignment stage follows this approach, using a multilingual text encoder instead.

\paragraph{Compact visual retrievers.} Several recent retrievers stay under 1B parameters: the late-interaction models ColUltraVBERT \citep{colultravbert} and ColSmol-500M \citep{colsmol2025}, the multi-vector retrievers pplx-embed-v2-late-0.6b \citep{pplxembedv2} and Vultron Flash \citep{vultron2026}, and the single-vector EmbeddingGemma-2 \citep{embeddinggemma2}. We compare \modelname{} with all of these, and with larger retrievers, on retrieval quality and on encoding speed measured in one harness on the same hardware (Section~\ref{subsec:speed}).

\paragraph{Late interaction and its cost.} ColBERT \citep{khattab2020colbert} scores a query against a document by summing, over query tokens, the maximum similarity to any document token (MaxSim). PLAID \citep{santhanam2022plaid} compresses and prunes this search, and MUVERA \citep{dhulipala2024muvera} reduces it to single-vector maximum inner-product search over fixed-dimensional encodings followed by exact reranking. Token pooling \citep{clavie2024tokenpooling} clusters document vectors to shrink the index, and MetaEmbed \citep{metaembed2025} learns a small set of summary vectors whose count can be chosen at test time. Matryoshka representation learning \citep{kusupati2024matryoshkarepresentationlearning} does the same for a single vector, training its leading dimensions to work as a shorter embedding. PLAID and MUVERA make search over a stored multi-vector index cheaper; we instead let the model's own single vector shortlist pages and either store or recompute the multi-vectors of the shortlist only, so a deployment can drop the multi-vector index altogether.

\paragraph{Granite Embedding.} The Granite Embedding R2 models \citep{awasthy2025graniteembeddingr2models,awasthy2026graniteembeddingmultilingualr2} are ModernBERT-based \citep{warner2024modernbert} text retrievers trained on governance-reviewed data. \modelname{} uses \textmodelname{} as its language backbone and reuses the R2 recipe's emphasis on knowledge distillation and model merging.

\section{Model Architecture}
\label{sec:architecture}

\begin{table}[h]
    \centering
    \small
    \begin{tabular}{llr}
    \toprule
        Component & Configuration & Params \\
    \midrule
        Vision encoder & SigLIP2 so400m NaFlex; 27 layers, hidden 1152; 576--2{,}880 patches of $16\times16$ px & 417.1M \\
        Connector & window Q-Former; 1 layer, 18 heads, $4\times4$ windows, hidden 1152 & 19.4M \\
        Vocabulary embeddings & $262{,}192\times768$ & 201.4M \\
        Text encoder & granite-embedding-311m-multilingual-r2; 22 layers, hidden 768 & 110.3M \\
        Single-vector head & \texttt{[CLS]} token, 768-d (Matryoshka: 128--768) & 0 \\
        Multi-vector head & linear projection $768\to128$, one vector per token & 0.1M \\
    \midrule
        Total &  & 748.3M \\
    \bottomrule
    \end{tabular}
    \caption{Architecture of \modelname{}, with parameter counts read from the checkpoint (748.3M parameters, 1.50\,GB in bf16). The vision encoder and the vocabulary table hold 83\% of the weights.}
    \label{tab:architecture}
\end{table}

\modelname{} is a vision-language encoder: visual tokens are processed alongside text tokens with full bidirectional attention (Figure~\ref{fig:architecture}, Table~\ref{tab:architecture}).

\paragraph{Vision encoder.} Pages are encoded by SigLIP2 so400m NaFlex \citep{tschannen2025siglip2} without its pooling head, initialized from the vision encoder of granite-vision-5.0-4b \citep{granitevision5}. Instead of resizing a page to a fixed grid of tiles, NaFlex splits it into patches at close to its native aspect ratio, so at inference a page is read at about 0.15 to 0.74 megapixels.

\paragraph{Connector.} A single-layer window Q-Former \citep{li2023blip2} compresses each $4\times4$ window of patch features, which reduces the number of visual tokens by $4\times$, and projects them into the text encoder's input space. A learned 2D position embedding and a newline embedding at the end of each row of patches help the text encoder recover the page layout. A page therefore reaches the text encoder as 144 to 720 visual tokens, about 2.5 times fewer than for the strongest compared model under 1B parameters, which keeps the model fast and its index small (Section~\ref{subsec:speed}).

\paragraph{Text encoder.} The backbone is \textmodelname{} \citep{awasthy2026graniteembeddingmultilingualr2}, a ModernBERT encoder \citep{warner2024modernbert} that alternates global and local attention. The visual tokens replace the image-placeholder tokens of the input sequence.

\paragraph{Output heads.} Both heads read the encoder's final hidden states:
\begin{itemize}
    \item \textbf{Single vector:} the L2-normalized final hidden state of the \texttt{[CLS]} token, with no parameters of its own. It is trained with Matryoshka representation learning \citep{kusupati2024matryoshkarepresentationlearning}, so it can be truncated to 128, 256, 384 or 512 dimensions.
    \item \textbf{Multi-vector:} a bias-free linear projection applied to every token, followed by L2 normalization. Queries and pages are scored with MaxSim:
    \[
        s_{\mathrm{MV}}(q, d) = \frac{1}{|q|}\sum_{i=1}^{|q|} \max_{j} \; \mathbf{q}_i^{\top} \mathbf{d}_j .
    \]
    Because the model is trained with a pooling-aware regularizer (Section~\ref{sec:kd}), the page vectors can be reduced at index time with hierarchical token pooling \citep{clavie2024tokenpooling} to a chosen number of vectors per page with little loss of quality (Section~\ref{sec:search}).
\end{itemize}

\section{Training}
\label{sec:training}

Training proceeds in two stages (Figure~\ref{fig:architecture}): modality alignment teaches the text encoder to read visual tokens, and knowledge distillation turns the aligned encoder into a retriever.

\subsection{Training Data}
\label{sec:training-data}

We curate a mix of public, synthetic and in-house data for the two training stages; Appendix~\ref{app:data} lists every dataset and its size.

\paragraph{Modality alignment} (6.3M examples) uses the commercially licensed subsets of The Cauldron \citep{laurencon2024idefics2}, Docmatix \citep{laurencon2024idefics3}, Molmo2-SynMultiImageQA \citep{clark2026molmo2} and question--answer pairs for the DocFM collection \citep{team2025granite}, followed by question--answer pairs and long descriptions of FinePDFs pages \citep{kydlicek2025finepdfs} in 22 languages.

\paragraph{Teacher training and distillation} draw on the same multimodal corpora, in different mixes. Distillation uses 7.3M multimodal training rows over 4.0M distinct pages and images, and adds text triples:
\begin{itemize}
    \item \textbf{Document retrieval pairs:} query--page pairs from document collections with mined hard negatives, including translated queries and documents, and queries derived from chain-of-thought VQA data.
    \item \textbf{Charts and tables:} chart question answering and reasoning data, including ChartNet \citep{chartnet2026} and its translations into six languages, chart captions, and financial tables.
    \item \textbf{Synthetic multilingual VQA:} queries for scientific papers, reports, financial documents and code images, written from a description of each image following ViDoRe v3 \citep{vidorev3} and translated from English into five languages.
    \item \textbf{Multilingual page retrieval:} MIRACL passages \citep{zhang-etal-2023-miracl} rendered in 16 languages, Wikimedia image captions in 10 languages, multi-page PDFs in 6 languages whose queries can span several similar pages, and FinePDFs pages in 22 languages, queried both with generated questions and with their alignment-stage descriptions.
    \item \textbf{Text retrieval data (distillation only):} the multilingual text triples of \citet{awasthy2026graniteembeddingmultilingualr2}, which preserve text retrieval.
\end{itemize}
Teacher training samples datasets with dataset weight temperature $T=2$ (meaning dataset sampling weights are set to the square-root of the length of the dataset). Distillation starts from the same temperature sampling and then balances text and multimodal data, with text at about 7\% of batches.

Synthetic queries, answers and descriptions are generated and verified with open-weight models (Gemma-4-26B-it, Gemma-4-31B-it and Ministral-3-3B-Instruct). All data undergoes a clearance process with technical, business, and governance review, capturing content description, intended use, data classification, licensing, usage restrictions, and assessment of sensitive information (e.g., personal information).

\subsection{Stage 1: Modality Alignment}
\label{sec:alignment}

Following ModernVBERT \citep{modernvbert2025}, we align the vision encoder with the text encoder using masked language modeling (MLM): half of the text tokens, in both questions and answers, are masked and reconstructed from the image and the remaining text. Because \textmodelname{} is already a strong text retriever, we add two \texttt{[CLS]} distillation losses with the frozen \textmodelname{} $f_{\theta_0}$ as the teacher. Let $f_\theta$ be the model being trained, $[\cdot]_{\mathrm{cls}}$ the \texttt{[CLS]} output, $\operatorname{sg}(\cdot)$ stop-gradient, and $z^{\mathrm{ref}} = \operatorname{sg}([f_{\theta_0}(x^{\mathrm{txt}})]_{\mathrm{cls}})$ the frozen model's embedding of a sample's text-only view. A \emph{text-view} loss keeps the text embedding in the original retrieval space, and an \emph{image-view} loss trains the embedding of the image-only view toward that of its text, after centering both by the batch mean $\mu$ of $z^{\mathrm{ref}}$:
\begin{equation}
  \mathcal{L}_{\mathrm{txt}} = 1 - \cos\!\big([f_\theta(x^{\mathrm{txt}})]_{\mathrm{cls}},\, z^{\mathrm{ref}}\big),
  \qquad
  \mathcal{L}_{\mathrm{img}} = 1 - \cos\!\big([f_\theta(x^{\mathrm{img}})]_{\mathrm{cls}} - \mu,\; z^{\mathrm{ref}} - \mu\big),
\end{equation}
\begin{equation}
  \mathcal{L}_{\mathrm{align}} = \mathcal{L}_{\mathrm{mlm}} + 0.5\,\mathcal{L}_{\mathrm{txt}} + \lambda\,\mathcal{L}_{\mathrm{img}},
  \qquad
  \lambda = \begin{cases} 0 & \text{step 1},\\ 1 & \text{steps 2 and 3}, \end{cases}
  \label{eq:align-loss}
\end{equation}
with all losses averaged over the batch.

Alignment runs in three steps, each initialized from the previous one. Step 1 trains with MLM and text-view distillation at up to 2{,}880 patches per image. Step 2 adds image-view distillation and raises the budget to 5{,}760 patches, on the same data: as with the high-resolution cooldown of ModernVBERT \citep{modernvbert2025}, this phase trains on higher-resolution input to improve accuracy, but without lowering the learning rate. Step 3 extends the model to more languages, especially low-resource ones such as Odia and Punjabi, by training on FinePDFs pages in 22 languages at up to 2{,}880 patches. Throughout, rank-32 DoRA adapters \citep{liu2024dora} on the text encoder and the connector are trained; step 3 also adds DoRA adapters on the vision encoder and trains its patch embeddings. Hyperparameters are in Table~\ref{tab:alignment-hparams} (Appendix~\ref{app:hyperparameters}).

\subsection{Stage 2: Knowledge Distillation}
\label{sec:kd}

\paragraph{Teachers.}\label{sec:teacher} Multimodal inputs are supervised by a multi-vector retriever built on Granite Vision 4.1 (4B parameters) \citep{granitevision2025} and trained with the ColPali recipe \citep{faysse2025colpali}. It is a merge of 11 checkpoints by linear interpolation \citep{wortsman2022soups} and averages 0.769 nDCG on the eight visual document retrieval benchmarks of Section~\ref{sec:evaluation}. Text-only inputs are supervised by the 8B single-vector teacher built on Granite 4.1 from \citet{awasthy2026graniteembeddingmultilingualr2}.

\paragraph{Distillation.} We distill from both teachers into both output heads at once. Starting from the aligned model, we run a short text warmup (a round of text-only distillation on triplet data with the text teacher), and then distill on the full mix. Each batch comes from a single dataset and is supervised by the teacher for its modality.


For a batch of $n$ queries, each query $q_i$ is scored against a candidate set $\mathcal{D}_i$ containing all in-batch negatives (gathered across devices) and one mined hard negative per query. Teacher scores $s^{t}(q_i, d)$ are MaxSim scores for the multimodal teacher and cosine similarities for the text teacher, computed online. For any scoring function $s$, we define the distribution over $\mathcal{D}_i$ at temperature $\tau = 0.02$ and the distillation loss against the teacher as
\[
    p_{i}[s](d) = \frac{\exp\left(s(q_i,d)/\tau\right)}{\sum_{d' \in \mathcal{D}_i} \exp\left(s(q_i,d')/\tau\right)},
    \qquad
    \mathcal{L}_{\mathrm{KD}}[s] = -\frac{1}{n}\sum_{i=1}^{n}\sum_{d \in \mathcal{D}_i} p_{i}[s^{t}](d)\, \log p_{i}[s](d).
\]
Both student heads are trained against the same teacher distribution $p_i[s^t]$, so they learn consistent rankings. This target is always computed from the full, uncompressed teacher representations, also when the student's representations are pooled or truncated below.

\paragraph{Multi-vector head with hierarchical pooling.} To make the multi-vector representation robust to index-time compression, we add a hierarchical pooling regularizer \citep{clavie2024tokenpooling}. For a budget of $r$ vectors, a document's token embeddings, except the leading \texttt{[CLS]}, are clustered into $r$ groups with Ward agglomerative clustering on cosine distance, and each group is replaced by its re-normalized centroid $\tilde{\mathbf{e}}_k = \bar{\mathbf{e}}_k / \lVert \bar{\mathbf{e}}_k \rVert_2$, giving the pooled document $d^{(r)}$. Documents with at most $r$ tokens are left unchanged, and queries are never pooled. The cluster assignment is treated as fixed in the backward pass (a straight-through estimator), so gradients flow through the centroids into the token embeddings. With $s^{(r)}_{\mathrm{MV}}(q, d) = s_{\mathrm{MV}}(q, d^{(r)})$,
\[
    \mathcal{L}_{\mathrm{MV}} = \mathcal{L}_{\mathrm{KD}}\!\left[s_{\mathrm{MV}}\right] + \lambda_{\mathrm{pool}}\, \mathcal{L}_{\mathrm{KD}}\!\left[s^{(r_t)}_{\mathrm{MV}}\right],
\]
where the budget $r_t$ is drawn uniformly at each step from $\mathcal{R} = \{32, 64, 128, 300\}$, with the same seed on all devices, and $\lambda_{\mathrm{pool}} = 0.5$. The student thus learns to reproduce the teacher's ranking even after its documents are compressed to $r$ vectors.

\paragraph{Single-vector head with Matryoshka representation learning.} For a dimension $m$, let $\mathbf{v}^{(m)}$ be the first $m$ coordinates of the single-vector embedding, re-normalized to unit length, and $s^{(m)}_{\mathrm{SV}}(q,d) = \cos\big(\mathbf{v}^{(m)}_q, \mathbf{v}^{(m)}_d\big)$. The Matryoshka objective \citep{kusupati2024matryoshkarepresentationlearning} sums the distillation loss over truncation levels:
\[
    \mathcal{L}_{\mathrm{SV}} = \sum_{m \in \mathcal{M}} \mathcal{L}_{\mathrm{KD}}\!\left[s^{(m)}_{\mathrm{SV}}\right], \qquad \mathcal{M} = \{128, 256, 384, 512, 768\}.
\]

\paragraph{Objective and trained parameters.} The full objective is $\mathcal{L} = 2\,\mathcal{L}_{\mathrm{MV}} + \mathcal{L}_{\mathrm{SV}}$, which weights the multi-vector head more to train strong multi-vector representations without losing single-vector quality. The vision encoder, connector and base text-encoder weights are frozen; we train rank-32 LoRA adapters \citep{hu2022lora} on the text encoder's attention and MLP projections, together with the multi-vector projection. Hyperparameters are in Table~\ref{tab:hyperparameters} (Appendix~\ref{app:hyperparameters}).

\subsection{Preserving Text Retrieval}
\label{sec:merge}

Multimodal training moves the text encoder away from its text-retrieval optimum. To recover text retrieval, we merge the text-encoder weights of the distilled student with those of \textmodelname{} (R2) by linear interpolation, $\theta = (1-\alpha)\,\theta_{\mathrm{MM}} + \alpha\,\theta_{\mathrm{R2}}$, which scales down the student's multimodal task vector \citep{ilharco2023taskarithmetic}; only the attention and MLP matrices differ between the two models. We chose $\alpha$ on an earlier checkpoint trained with the same recipe, using a reduced Natural Questions dev-test pool \citep{kwiatkowski-etal-2019-natural-questions} for text and ViDoRe v1 for images (Table~\ref{tab:merge-alpha}, Appendix~\ref{app:hyperparameters}). Larger $\alpha$ trades visual for text retrieval, and $\alpha = 0.1$ is the only weight we tested at which the multi-vector text gain exceeds the image cost (+0.014 NQ nDCG@10 against $-$0.006 ViDoRe v1 nDCG@5); it also raises single-vector text retrieval by 0.017. Spherical interpolation and per-layer schedules gave no further gains.



\section{Evaluation}
\label{sec:evaluation}

\subsection{Setup}
We evaluate visual document retrieval (text queries, page-image documents) on eight benchmarks: ViDoRe v1 and v2 \citep{faysse2025colpali}, ViDoRe v3 \citep{vidorev3}, Real-MM-RAG \citep{realmmrag2025}, Jina-VDR \citep{gunther2025jinav4}, M3DR \citep{m3dr2025}, VisRAG \citep{visrag} and VisDocOOD, the four out-of-domain visual document tasks of MMEB-V2 \citep{visdocood} (ViDoSeek and MMLongBench-Doc, each at page and document level). For text retrieval we use the retrieval tasks of the MTEB English and multilingual benchmarks \citep{muennighoff2022mteb,enevoldsen2025mmtebmassivemultilingualtext}. We report nDCG@5 for visual retrieval (nDCG@10 for ViDoRe v3) and nDCG@10 for text. 

Our main baselines are the five retrievers under 1B parameters of Section~\ref{sec:related}: pplx-embed-v2-late-0.6b, VultronRetrieverFlash-Qwen3.5-0.8B (Vultron Flash below), ColUltraVBERT, ColSmol-500M and EmbeddingGemma-2. For the speed--quality comparison (Section~\ref{subsec:speed}) we add 15 larger open retrievers, for 20 in total \citep{nemotron35embed,jinav5omni2026,qwen3vlembedding,topkembed,argus2026,colqwen25,gunther2025jinav4,evie,webaicolvec,vultroncore,nemotroncolembedv2,colnomic2025}, listed in Table~\ref{tab:speed} (Appendix~\ref{app:speed}).

\paragraph{Evaluation protocol.} We measure encoding speed as page images encoded per second over the 14 tasks of ViDoRe v1 and v2 (20{,}280 pages), on one NVIDIA H200 per model, in bf16 at batch size 32, after an untimed warm-up task. This includes each model's own image preprocessing, so it measures end-to-end encoding rather than GPU compute alone. Each model runs with the transformers version its implementation requires. Scores are computed in bf16, where a page's embedding varies slightly with the other pages in its batch; between runs this can change a benchmark average by about $\pm0.0015$, so differences of a few thousandths are within noise. Baseline scores on ViDoRe v1--v3 and Jina-VDR are taken from the MTEB leaderboard where available; all others, and all scores of jina-embeddings-v4\footnote{With its default cap of 602{,}112 pixels per image; without it, the largest ViDoRe v1 scans need a single 69\,GiB allocation.} and EmbeddingGemma-2\footnote{Its MTEB implementation frames most page images with the query template instead of the document prompt from its paper (\texttt{title: none | text: <image>}). 
We use the paper's prompt, which raises its eight-benchmark average from 0.567 to 0.606.}, are from our own runs. 

\subsection{Retrieval Quality}
\label{subsec:visual-results}

\begin{table}[t]
    \centering
    {\renewcommand{\arraystretch}{1.1}
    \begin{adjustbox}{width=\textwidth}
    \begin{tabular}{l|r|r|cccccccc|c}
    \toprule
        Model & Params & Dims & ViDoRe v1 & ViDoRe v2 & ViDoRe v3 & Real-MM & Jina-VDR & M3DR & VisRAG & VisDocOOD & Avg. \\
        & & & (10) & (4) & (8) & (4) & (43) & (23) & (6) & (4) & \\
        \hline
        \emph{Multi-vector models} & & & & & &  &  &  &  &  & \\
        pplx-embed-v2-late-0.6b & 594M & 128 & \textbf{0.897} & \textbf{0.652} & \textbf{0.614} & \textbf{0.897} & \textbf{0.782} & \textbf{0.602} & 0.866 & \textbf{0.715} & \textbf{0.753} \\
        \modelname{} (multi-vector) & 748M & 128 & 0.864 & 0.629 & 0.561 & 0.859 & 0.747 & 0.548 & 0.857 & 0.704 & 0.721 \\
        Vultron Flash & 853M & 320 & 0.883 & 0.604 & 0.565 & 0.842 & 0.741 & 0.479 & \textbf{0.884} & 0.685 & 0.710 \\
        ColUltraVBERT & 255M & 768 & 0.863 & 0.546 & 0.486 & 0.855 & 0.664 & 0.298 & 0.862 & 0.693 & 0.658 \\
        ColSmol-500M & 507M & 128 & 0.824 & 0.429 & 0.361 & 0.731 & 0.592 & 0.220 & 0.825 & 0.666 & 0.581 \\
        \hline
        \emph{Single-vector models} & & & & & &  &  &  &  &  & \\
        EmbeddingGemma-2$^\ddagger$ & 744M & 768 & 0.728 & 0.522 & 0.445 & 0.615 & 0.652 & 0.496 & 0.732 & 0.660 & 0.606 \\
        \modelname{} (single vector) & 748M & 768 & 0.745 & 0.542 & 0.443 & 0.648 & 0.635 & 0.422 & 0.738 & 0.638 & 0.601 \\
        \modelname{} (single vector, 384-d) & 748M & 384 & 0.742 & 0.537 & 0.438 & 0.643 & 0.631 & 0.419 & 0.729 & 0.636 & 0.597 \\
        \hline
        \textcolor{gray}{Teacher (Granite Vision 4.1)} & \textcolor{gray}{4B} & \textcolor{gray}{128} & \textcolor{gray}{0.903} & \textcolor{gray}{0.692} & \textcolor{gray}{0.627} & \textcolor{gray}{0.903} & \textcolor{gray}{0.805} & \textcolor{gray}{0.596} & \textcolor{gray}{0.900} & \textcolor{gray}{0.726} & \textcolor{gray}{0.769} \\
    \bottomrule
    \end{tabular}
    \end{adjustbox}}
    \caption{Visual document retrieval against the other models under 1B parameters, and the teacher. nDCG@5, except ViDoRe v3 (nDCG@10); the number of tasks is in parentheses. Dims is the dimension per token for multi-vector models. Bold marks the best score in each column among the models under 1B. $^\ddagger$With the document prompt from its paper and 280 soft tokens per page, the checkpoint default.}
    \label{tab:visual-results}
\end{table}

Table~\ref{tab:visual-results} compares \modelname{} with the other retrievers under 1B parameters and with its 4B teacher. Its multi-vector head averages 0.721, second among these models: only pplx-embed-v2-late-0.6b scores higher (0.753), at half the encoding speed (Section~\ref{subsec:speed}), and Vultron Flash, ColUltraVBERT and ColSmol-500M score 0.011 to 0.140 lower. Against Vultron Flash, the closest model in size, it is ahead on five of the eight benchmarks (ViDoRe v2, Real-MM-RAG, Jina-VDR, M3DR and VisDocOOD) with 2.5$\times$ narrower token vectors (128 dimensions against 320). It keeps 94\% of the teacher's average (0.721 against 0.769) with under a fifth of its parameters.

The single-vector head averages 0.601, 0.005 below EmbeddingGemma-2 and ahead of it on four of the eight benchmarks; truncated to 384 dimensions, it loses 0.004 on average, and at most 0.009 on any benchmark, for half the storage.

On the multilingual benchmarks, the multi-vector head is second among the models under 1B parameters on both M3DR (0.548, against 0.602 for pplx-embed-v2-late-0.6b and 0.479 for Vultron Flash) and Jina-VDR (0.747, against 0.782 and 0.741), while the two smallest multi-vector models fall below 0.30 on M3DR. The single-vector head is weaker here, trailing EmbeddingGemma-2 on M3DR (0.422 against 0.496).

\subsection{Encoding Cost}
\label{subsec:speed}

Figure~\ref{fig:pareto} (data in Table~\ref{tab:speed}, Appendix~\ref{app:speed}) places \modelname{} among open visual retrievers by quality and encoding speed. Its multi-vector head encodes 35.3 pages/s, faster than every other model we compared, at an average of 0.721, so it lies on the Pareto front: every model that scores higher is slower. At this rate one H200 encodes a million pages in about 8 hours, against 16 to 107 hours for the models that score higher. Among the models under 1B parameters, only pplx-embed-v2-late-0.6b scores higher (0.753), at 17.4 pages/s, half the speed; it passes about 1{,}500 tokens per ViDoRe page through its language model, 2.5 times the 590 of \modelname{}.\footnote{The optional fused kernels for its linear-attention layers were not available in our environment, so it ran without them. We expect this to matter little: in this harness page decoding dominates, and FlashAttention-2 sped up page encoding of \modelname{} by only 4\%.} Vultron Flash is both slower (17.1 pages/s) and lower (0.710). The larger models that score higher have 2.2 to 9B parameters and run at 0.07 to 0.35 of its speed, and Argus-ColQwen3.5-2B matches it (0.721) at 9.4 pages/s. EmbeddingGemma-2, the next-fastest model, encodes 28.2 pages/s and scores 0.115 lower.


As a standalone retriever, the single-vector head (0.601, also 35.3 pages/s) is 25\% faster than EmbeddingGemma-2 and 0.005 lower, so it is on the single-vector Pareto front (Figure~\ref{fig:pareto-sv}). It is meant as the first stage of hybrid search, where it comes from the same forward pass as the multi-vectors that rerank its candidates (Section~\ref{sec:search}); Appendix~\ref{app:vs-gemma} compares the two models further.


\paragraph{Where the speed comes from.}\label{subsec:inference-speedup}

\begin{table}[h]
    \centering
    \small
    \begin{tabular}{l|r|r|l}
    \toprule
        Configuration & Pages/s $\uparrow$ & Speed-up & Encoder output \\
        \hline
        Serial CPU preprocessing, standard attention & 26.3 & 1.00$\times$ & reference \\
        + threaded CPU preprocessing & 30.6 & 1.16$\times$ & bit-identical \\
        + GPU preprocessing & 32.0 & 1.22$\times$ & resize rounding \\
        + FlashAttention-2 & 33.4 & 1.27$\times$ & bf16 kernel differences \\
        Release measurement (Table~\ref{tab:speed}) & \textbf{35.3} & \textbf{1.34$\times$} & identical scores \\
    \bottomrule
    \end{tabular}
    \caption{Changes to the inference path, each added to the row above. Single-vector head; the multi-vector head runs at the same speed, since both come from one encoding pass. Pages/s on one H200 over ViDoRe v1 and v2, as in Table~\ref{tab:speed}. The last row is the measurement of Table~\ref{tab:speed}, from a later run with a newer mteb version. Differences of about 2\% are within node-to-node variation.}
    \label{tab:speedup}
\end{table}

Every model in Table~\ref{tab:speed} runs with its released preprocessing, and for \modelname{} that includes optimizations to the image processor and attention that ship with the model; none of them changes the weights or the architecture. We first measured where the time goes on an H200. The MTEB harness decodes every stored page inside the timed loop (about 27\,ms per page on DocVQA), and the original image processor resized, normalized and split each page into patches serially on the CPU. Pages also vary in size, from 576 to 2{,}880 patches: with standard attention a 32-page batch is padded to its largest page, which adds 12.4\% more patches over ViDoRe v1 and v2, and this padding passes through all 27 layers of the vision tower. We made three changes, whose effect is shown in Table~\ref{tab:speedup}:
\begin{itemize}
    \item \textbf{Threaded preprocessing.} The image processor resizes, normalizes and patchifies the pages of a batch on a thread pool of up to eight workers, speeding up the preprocessing with no change to the output.
    \item \textbf{GPU preprocessing.} When a GPU is available, the same pipeline runs there; the CPU path remains for CPU-only machines. The output differs from the CPU path only by resize rounding (mean absolute difference $6\cdot10^{-4}$ on the $[-1,1]$ pixel scale).
    \item \textbf{FlashAttention-2.} Under FlashAttention-2 \citep{dao2023flashattention2}, the vision tower packs the pages of a batch into one variable-length sequence, so no padding is computed; the text encoder uses it as well. Without FlashAttention-2 installed, the model falls back to PyTorch's scaled dot-product attention.
\end{itemize}

\paragraph{Effect.} Together, the changes make \modelname{} 1.27$\times$ faster (26.3 to 33.4 pages/s), and 1.34$\times$ in the final release measurement (35.3 pages/s), without changing its quality: on five dev sets GPU and CPU preprocessing differ by at most 0.0008 macro nDCG@5, within 1.6 standard errors, and with GPU preprocessing and FlashAttention-2 together the eight-benchmark average moves by less than 0.001, and no benchmark by more than 0.003. FlashAttention-2 matters much more when page decoding is not timed: on an RTX 5090 with pages decoded in advance it raises throughput from 23.9 to 39.7 pages/s (1.66$\times$), against 1.04$\times$ in the H200 harness. EmbeddingGemma-2 gains almost nothing from it under the same protocol (37.5 to 40.1 pages/s, nearly all on two noisy tasks): its text model's full-attention layers have 512-dimensional heads, beyond FlashAttention-2's limit of 256, and its vision tower has fixed-size inputs with no padding to remove. With decoding outside the timed loop the two models therefore encode at about the same rate, and \modelname's lead in Table~\ref{tab:speed} comes from its preprocessing path. Appendix~\ref{app:speedup-extra} describes further speed-ups outside the model, such as overlapping page decoding with encoding and an FP8 vision tower, which made an earlier checkpoint 2.4$\times$ faster in this harness.

\subsection{Serving Cost and Index Size}
\label{sec:search}

\begin{table}[h]
    \centering
    \small
    \begin{adjustbox}{max width=\textwidth}
    \begin{tabular}{l|r|ccc|r}
    \toprule
        Configuration & Stored per page & ViDoRe v1 & ViDoRe v2 & Avg. & Latency / query \\
        \hline
        Single vector & 1.5 KiB & 0.745 & 0.542 & 0.644 & 15.1\,ms \\
        Hybrid, stored multi-vectors ($k=100$) & 149 KiB (75 int8) & 0.861 & 0.630 & 0.745 & 16.5\,ms$^\dagger$ \\
        Hybrid, recomputed multi-vectors ($k=10$) & 1.5 KiB + image & 0.827 & 0.616 & 0.721 & 223 / 383\,ms \\
        Hybrid, recomputed multi-vectors ($k=20$) & 1.5 KiB + image & 0.838 & 0.621 & 0.730 & 379 / 739\,ms \\
        Hybrid, recomputed multi-vectors ($k=100$) & 1.5 KiB + image & 0.852 & 0.629 & 0.740 & 1.50\,s \\
        Multi-vector, exhaustive & 147 KiB & 0.864 & 0.629 & 0.746 & 15.2\,ms$^\ast$ \\
    \bottomrule
    \end{tabular}
    \end{adjustbox}
    \caption{Search configurations of the released model, nDCG@5 averaged over 10 (ViDoRe v1) and 4 (ViDoRe v2) datasets. Measured on an H100 in bf16, one query at a time; latency includes query encoding. Storage is at fp16. Recomputed reranking keeps only the single vector and the page image, which is stored anyway; its latencies are with the shortlisted pages encoded as one batch / one at a time ($k=100$: batched only), and its scores are on a fixed sample of 100 queries per dataset, so they differ slightly from the other rows. $^\dagger$19.7\,ms when the multi-vectors are read from host memory. $^\ast$ViDoRe corpora have at most 1{,}600 pages.}
    \label{tab:search}
\end{table}

Because one model produces both embeddings, a deployment can trade accuracy against cost at search time:
\begin{itemize}
    \item \textbf{Single-vector search:} one 768-dimensional vector per page, served by any approximate nearest-neighbor index.
    \item \textbf{Multi-vector search:} 128-dimensional vectors for every visual token of a page, scored with MaxSim.
    \item \textbf{Hybrid search:} retrieve the top-$k$ pages with the single-vector index, then rerank them with exact MaxSim, with multi-vectors either stored or recomputed for the $k$ candidates at query time.
\end{itemize}

We propose hybrid search, in either variant, as an alternative to exhaustive multi-vector search; Table~\ref{tab:search} compares both with single-vector and exhaustive multi-vector search on ViDoRe v1 and v2. With stored multi-vectors, reranking 100 single-vector candidates adds 0.116 nDCG@5 on ViDoRe v1 for about 1.5\,ms more per query, and is on average only 0.001 below exhaustive multi-vector search; its cost is storage, since a page's multi-vectors take about 100$\times$ the space of its single vector (147\,KiB against 1.5\,KiB at fp16, for about 590 vectors per page). With recomputed multi-vectors, the index keeps only the single vector and the page image, which is stored anyway; re-encoding the shortlist costs 15--20\,ms per page in a batch (about 35\,ms one at a time), so 0.2--0.4\,s per query at $k=10$, and reaches 0.721 at $k=10$ and 0.740 at $k=100$ on a fixed sample of 100 queries per dataset. On the same queries it ranks as stored reranking does: identically when pages are re-encoded one at a time, and within 0.003 in a batch, which changes their bf16 embeddings slightly. Exhaustive multi-vector search is fast on ViDoRe only because each corpus has at most 1{,}600 pages; its cost grows linearly with corpus size, and at one million pages its vectors alone take about 150\,GB.

\paragraph{Single vector or fixed-dimensional encoding as the first stage.} MUVERA \citep{dhulipala2024muvera} generates candidates for multi-vector reranking from fixed-dimensional encodings (FDEs) of the token sets. The single vector of \modelname{} does this job better with less storage: on an earlier checkpoint, with the same exact-MaxSim reranking, a shortlist from the 768-dimensional single vector (3\,KB per page at fp32) beat one from a 10{,}240-dimensional FDE (40\,KB per page, 13$\times$ the storage) at every depth from 5 to 200 and on all 22 tasks of ViDoRe v1--v3. The gain was largest at shallow depths, 0.08 to 0.11 nDCG@5 at a depth of 5, and the two converged at the deepest shortlists. A model with a single-vector head therefore needs no separate FDE index.

\paragraph{Token pooling and int8.} Because the model is trained with a pooling-aware regularizer (Section~\ref{sec:kd}), hierarchical token pooling of the page vectors at index time \citep{clavie2024tokenpooling} costs little quality; queries are not pooled. Pooling to 256 vectors per page makes the index 56\% smaller while keeping 99.6\% of full quality, 128 vectors (78\% smaller) keep 98.8\%, and 32 vectors (94\% smaller) still keep 96.1\% (Table~\ref{tab:pooling}). Storing the page vectors in int8 (one byte per dimension, a fixed scale of 127 on the unit-length vectors) halves the index at no measurable cost: the average moves by $+0.0007$, and across all pooling budgets int8 differs significantly from fp16 on only 2 of 98 dataset--budget pairs, with opposite signs. The two savings therefore compose: int8 with 128 vectors per page makes the index 9.1$\times$ smaller than the unpooled fp16 index for a loss of 0.009, and with 32 vectors 36$\times$ smaller for 0.030.

\begin{table}[t]
    \centering
    \small
    \begin{tabular}{r|ccc|rrr}
    \toprule
        Vectors / page & ViDoRe v1 & ViDoRe v2 & Avg. & Retained & Index saved & At 1M pages \\
        \hline
        full (589) & 0.864 & 0.629 & 0.746 & -- & -- & 140 GiB \\
        300 & 0.861 & 0.624 & 0.743 & 99.6\% & 48.9\% & 72 GiB \\
        256 & 0.860 & 0.625 & 0.743 & 99.6\% & 56.4\% & 61 GiB \\
        128 & 0.855 & 0.618 & 0.737 & 98.8\% & 78.1\% & 31 GiB \\
        64 & 0.847 & 0.611 & 0.729 & 97.8\% & 89.0\% & 15.5 GiB \\
        32 & 0.833 & 0.601 & 0.717 & 96.1\% & 94.4\% & 7.9 GiB \\
        16 & 0.807 & 0.584 & 0.695 & 93.2\% & 97.1\% & 4.1 GiB \\
    \bottomrule
    \end{tabular}
    \caption{Hierarchical token pooling of the page multi-vectors of the released model, exhaustive MaxSim, nDCG@5 over 10 (ViDoRe v1) and 4 (ViDoRe v2) datasets. A page keeps its \texttt{[CLS]} vector and at most the stated number of pooled vectors; unpooled pages average 589 vectors. Retained is relative to the unpooled average; index sizes are at fp16 (128 dimensions). Pages encoded one at a time on an H100 in bf16; the unpooled row is the exhaustive row of Table~\ref{tab:search}.}
    \label{tab:pooling}
\end{table}

\subsection{Text Retrieval and Index Sharing}
\label{subsec:text-results}

\begin{table}[ht]
    \centering
    {\renewcommand{\arraystretch}{1.1}
    \begin{adjustbox}{max width=\textwidth}
    \begin{tabular}{l|r|l|cc}
    \toprule
        Model & Params & Type & MTEB (eng) & MTEB (multilingual) \\
        \hline
        pplx-embed-v2-late-0.6b \citep{pplxembedv2} & 594M & multi-vector & \textbf{0.567} & 0.632 \\
        EmbeddingGemma-2 \citep{embeddinggemma2} & 744M & single vector & 0.513 & 0.606 \\
        Vultron Flash \citep{vultron2026} & 853M & multi-vector & 0.467 & 0.532 \\
        granite-embedding-311m-multilingual-r2 (text only) & 311M & single vector & 0.526 & 0.652 \\
        \hline 
        \modelname{} & 748M & multi-vector & 0.509 & \textbf{0.656} \\
        \modelname{} & 748M & single vector & 0.478 & 0.629 \\
        \modelname{} & 748M & single vector, 384-d & 0.477 & 0.628 \\
        
    \bottomrule
    \end{tabular}
    \end{adjustbox}}
    \caption{Text retrieval, nDCG@10 averaged over the MTEB retrieval tasks (10 English, 18 multilingual; 25 distinct tasks). Bold marks the best score in each column; multi-vector models are scored with exhaustive MaxSim over the full corpus, and single-vector models with cosine similarity. Scores for Vultron Flash, pplx-embed-v2-late-0.6b and \modelname{} are our own runs; the others are from the MTEB leaderboard.}
    \label{tab:text-results}
\end{table}

\modelname{} remains a strong text retriever (Table~\ref{tab:text-results}). Its multi-vector head scores highest of these models on the multilingual suite (0.656), slightly above the single-vector \textmodelname{} (0.652), and trails it on English (0.509 against 0.526); its single-vector head trails it on both suites (0.629 and 0.478). Against the other models of its size, the multi-vector head is ahead of Vultron Flash on both suites, and the single-vector head is ahead of EmbeddingGemma-2 on the multilingual suite (0.629 against 0.606) and behind it on English (0.478 against 0.513); pplx-embed-v2-late-0.6b is clearly the strongest on English. Truncating the single vector to 384 dimensions costs only 0.001 on each suite, for half the storage. For text-only corpora, \textmodelname{} therefore remains the better choice, with well under half the parameters and one vector per passage, and the two can share an index: on the 25 retrieval tasks of the two MTEB suites, queries encoded by \modelname{} retrieve from the single vectors of \textmodelname{} at least as well as from its own (Appendix~\ref{app:text-compat}). An existing \textmodelname{} text index can therefore be kept as it is when page images are added.

\section{Conclusion}
\modelname{} shows that competitive visual document retrieval does not need a multi-billion-parameter model. A 748M-parameter encoder distilled from a 4B multi-vector teacher keeps 94\% of the teacher's quality (0.721 against 0.769 nDCG over eight benchmarks) and encodes 35.3 pages per second on one H200, the fastest of the retrievers we compared, which places it on the speed--quality Pareto front. For enterprise collections that speed matters as much as the last points of accuracy: one H200 indexes a million pages in about 8 hours, against 16 to 107 hours for the models that score higher.

Two design choices carry over beyond this model. First, one forward pass yields both a single vector and narrow multi-vectors: the single vector shortlists pages for the multi-vectors to rerank, which can be stored or recomputed for the shortlist only, so a deployment need not keep a multi-vector index at all. Both outputs can also be shrunk after training, by token pooling and Matryoshka truncation, so each deployment chooses its own point between cost and quality when it builds its index rather than when the model is trained. Second, adapting a frozen multilingual text encoder with LoRA and merging back toward it keeps much of what the text model already knew: the multi-vector head matches the single-vector \textmodelname{} on multilingual retrieval (0.656 against 0.652), and its queries can search \textmodelname{}'s single vectors, so an existing text index can take page images without being re-encoded.

\paragraph{Ethical Considerations and Limitations.} 
\modelname{} is optimized for visual document retrieval. The model is not trained to mitigate demographic biases or harmful visual-text associations, particularly when the model is applied outside its intended domain. This model is not intended for assessments of individuals. Users should evaluate retrieval behavior and potential representational harms in their specific application before deployment.

When the same document exists in several languages, \modelname{} finds the right page but not always its rendering in the query's script. We measured this on development pages of M3DR \citep{m3dr2025} rendered in 18 languages, with a release candidate. Each page's text finds a rendering of the same page first in 98--99\% of cases in every language, and the rendering in its own language in 97\% of cases when only that language's pages are searched. Across all languages, the text's own rendering comes first most reliably for scripts used by a single language in the pool, such as Thai, Gurmukhi, Japanese and Chinese (50--57\% of texts); Latin-script texts fall in between (34--48\%), most often losing to the English rendering, and Indic languages fail most: their queries retrieve the native page first in under 10\% of cases, and the English copy in 65--86\%. The teacher shows similar behavior. In practice the weakness is easy to mitigate when the language of each page is known, from document metadata or language identification: identifying the query's language and restricting the search to pages in that language puts the text's own rendering first in 97\% of cases, as above. Teaching the model itself to prefer the native rendering would need a contrastive objective that ties a text to the page in its own language.

\paragraph{Future work.} Directions include narrowing the remaining gap to the teacher and to pplx-embed-v2-late-0.6b, which scores 0.032 higher at half the speed, without giving up encoding speed, and aligning the two models in the other direction, so that queries encoded by \textmodelname{} can search a \modelname{} index on every task and retrieve page images from it (Appendix~\ref{app:text-compat}).

\bibliography{main,mm}

\begin{thebibliography}{63}
\providecommand{\natexlab}[1]{#1}
\providecommand{\url}[1]{\texttt{#1}}
\expandafter\ifx\csname urlstyle\endcsname\relax
  \providecommand{\doi}[1]{doi: #1}\else
  \providecommand{\doi}{doi: \begingroup \urlstyle{rm}\Url}\fi

\bibitem[Abdallah et~al.(2026)Abdallah, Abdalla, Ali, and Jatowt]{argus2026}
Abdelrahman Abdallah, Mahmoud Abdalla, Mohammed Ali, and Adam Jatowt.
\newblock {Argus-Retriever}: Vision-{LLM} late-interaction retrieval with
  region-aware query-conditioned {MoE} for visual document retrieval.
\newblock \url{https://huggingface.co/DataScience-UIBK/Argus-Colqwen3.5-2b-v0},
  2026.
\newblock URL \url{https://arxiv.org/abs/2606.04300}.

\bibitem[{athrael-soju}(2026)]{colultravbert}
{athrael-soju}.
\newblock {ColUltraVBERT}.
\newblock \url{https://huggingface.co/athrael-soju/ColUltraVBERT}, 2026.

\bibitem[Awasthy et~al.(2025)Awasthy, Trivedi, Li, Doshi, Bhat, P, Kumar, Yang,
  Iyer, Daniels, Murthy, Barker, Franz, Lee, Ward, Roukos, Cox, Lastras, Sen,
  and Florian]{awasthy2025graniteembeddingr2models}
Parul Awasthy, Aashka Trivedi, Yulong Li, Meet Doshi, Riyaz Bhat, Vignesh P,
  Vishwajeet Kumar, Yushu Yang, Bhavani Iyer, Abraham Daniels, Rudra Murthy,
  Ken Barker, Martin Franz, Madison Lee, Todd Ward, Salim Roukos, David Cox,
  Luis Lastras, Jaydeep Sen, and Radu Florian.
\newblock Granite embedding r2 models, 2025.
\newblock URL \url{https://arxiv.org/abs/2508.21085}.

\bibitem[Awasthy et~al.(2026)Awasthy, Trivedi, Yang, Barker, Li, Iyer, Franz,
  Bross, Doshi, P, Kumar, Ward, Daniels, Lee, Lastras, Sen, and
  Florian]{awasthy2026graniteembeddingmultilingualr2}
Parul Awasthy, Aashka Trivedi, Yushu Yang, Ken Barker, Yulong Li, Bhavani Iyer,
  Martin Franz, Juergen Bross, Meet Doshi, Vignesh P, Vishwajeet Kumar, Todd
  Ward, Abraham Daniels, Madison Lee, Luis Lastras, Jaydeep Sen, and Radu
  Florian.
\newblock Granite embedding multilingual r2 models, 2026.
\newblock URL \url{https://arxiv.org/abs/2605.13521}.

\bibitem[Chang et~al.(2022)Chang, Palzer, Li, Fosler-Lussier, and
  Xiao]{chang2022mapqa}
Shuaichen Chang, David Palzer, Jialin Li, Eric Fosler-Lussier, and Ningchuan
  Xiao.
\newblock {MapQA}: A dataset for question answering on choropleth maps.
\newblock In \emph{NeurIPS 2022 Workshop on Table Representation Learning},
  2022.
\newblock URL \url{https://openreview.net/forum?id=znKbVjeR0yI}.

\bibitem[Chen et~al.(2026)Chen, Liu, Luo, Wang, Yang, Wei, and Dou]{moca2025}
Haonan Chen, Hong Liu, Yuping Luo, Liang Wang, Nan Yang, Furu Wei, and Zhicheng
  Dou.
\newblock {MoCa}: Modality-aware continual pre-training makes better
  bidirectional multimodal embeddings.
\newblock In \emph{Proceedings of the 64th Annual Meeting of the Association
  for Computational Linguistics (Volume 1: Long Papers)}, pp.\  810--823, San
  Diego, California, United States, July 2026. Association for Computational
  Linguistics.
\newblock \doi{10.18653/v1/2026.acl-long.34}.
\newblock URL \url{https://aclanthology.org/2026.acl-long.34/}.

\bibitem[Clark et~al.(2026)Clark, Zhang, Ma, Park, Tripathi, Lee, Salehi, Ren,
  Kim, Yang, et~al.]{clark2026molmo2}
Christopher Clark, Jieyu Zhang, Zixian Ma, Jae~Sung Park, Rohun Tripathi,
  Sangho Lee, Mohammadreza Salehi, Jason Ren, Chris~Dongjoo Kim, Yinuo Yang,
  et~al.
\newblock Molmo2: Open weights and data for vision-language models with video
  understanding and grounding.
\newblock In \emph{Proceedings of the IEEE/CVF Conference on Computer Vision
  and Pattern Recognition}, pp.\  28652--28668, 2026.

\bibitem[Clavi{\'e} et~al.(2024)Clavi{\'e}, Chaffin, and
  Adams]{clavie2024tokenpooling}
Benjamin Clavi{\'e}, Antoine Chaffin, and Griffin Adams.
\newblock Reducing the footprint of multi-vector retrieval with minimal
  performance impact via token pooling, 2024.

\bibitem[Dao(2023)]{dao2023flashattention2}
Tri Dao.
\newblock Flashattention-2: Faster attention with better parallelism and work
  partitioning, 2023.

\bibitem[de~Souza P.~Moreira et~al.(2026)de~Souza P.~Moreira, Ak, Xu,
  Holworthy, Schifferer, Yu, Babakhin, Osmulski, Cai, Chesler, Liu, and
  Oldridge]{nemotroncolembedv2}
Gabriel de~Souza P.~Moreira, Ronay Ak, Mengyao Xu, Oliver Holworthy, Benedikt
  Schifferer, Zhiding Yu, Yauhen Babakhin, Radek Osmulski, Jiarui Cai, Ryan
  Chesler, Bo~Liu, and Even Oldridge.
\newblock {Nemotron ColEmbed V2}: Top-performing late interaction embedding
  models for visual document retrieval.
\newblock \url{https://huggingface.co/nvidia/nemotron-colembed-vl-4b-v2},
  \url{https://huggingface.co/nvidia/nemotron-colembed-vl-8b-v2}, 2026.
\newblock URL \url{https://arxiv.org/abs/2602.03992}.

\bibitem[Dhulipala et~al.(2024)Dhulipala, Hadian, Jayaram, Lee, and
  Mirrokni]{dhulipala2024muvera}
Laxman Dhulipala, Majid Hadian, Rajesh Jayaram, Jason Lee, and Vahab Mirrokni.
\newblock {MUVERA}: Multi-vector retrieval via fixed dimensional encodings.
\newblock In \emph{Advances in Neural Information Processing Systems}, 2024.

\bibitem[Enevoldsen et~al.(2025)Enevoldsen, Chung, Kerboua, Kardos, Mathur,
  Stap, Gala, Siblini, Krzemiński, Winata, Sturua, Utpala, Ciancone,
  Schaeffer, Sequeira, Misra, Dhakal, Rystrøm, Solomatin, Ömer Çağatan,
  Kundu, Bernstorff, Xiao, Sukhlecha, Pahwa, Poświata, GV, Ashraf, Auras,
  Plüster, Harries, Magne, Mohr, Hendriksen, Zhu, Gisserot-Boukhlef, Aarsen,
  Kostkan, Wojtasik, Lee, Šuppa, Zhang, Rocca, Hamdy, Michail, Yang, Faysse,
  Vatolin, Thakur, Dey, Vasani, Chitale, Tedeschi, Tai, Snegirev, Günther,
  Xia, Shi, Lù, Clive, Krishnakumar, Maksimova, Wehrli, Tikhonova, Panchal,
  Abramov, Ostendorff, Liu, Clematide, Miranda, Fenogenova, Song, Safi, Li,
  Borghini, Cassano, Su, Lin, Yen, Hansen, Hooker, Xiao, Adlakha, Weller,
  Reddy, and Muennighoff]{enevoldsen2025mmtebmassivemultilingualtext}
Kenneth Enevoldsen, Isaac Chung, Imene Kerboua, Márton Kardos, Ashwin Mathur,
  David Stap, Jay Gala, Wissam Siblini, Dominik Krzemiński, Genta~Indra
  Winata, Saba Sturua, Saiteja Utpala, Mathieu Ciancone, Marion Schaeffer,
  Gabriel Sequeira, Diganta Misra, Shreeya Dhakal, Jonathan Rystrøm, Roman
  Solomatin, Ömer Çağatan, Akash Kundu, Martin Bernstorff, Shitao Xiao,
  Akshita Sukhlecha, Bhavish Pahwa, Rafał Poświata, Kranthi~Kiran GV, Shawon
  Ashraf, Daniel Auras, Björn Plüster, Jan~Philipp Harries, Loïc Magne,
  Isabelle Mohr, Mariya Hendriksen, Dawei Zhu, Hippolyte Gisserot-Boukhlef, Tom
  Aarsen, Jan Kostkan, Konrad Wojtasik, Taemin Lee, Marek Šuppa, Crystina
  Zhang, Roberta Rocca, Mohammed Hamdy, Andrianos Michail, John Yang, Manuel
  Faysse, Aleksei Vatolin, Nandan Thakur, Manan Dey, Dipam Vasani, Pranjal
  Chitale, Simone Tedeschi, Nguyen Tai, Artem Snegirev, Michael Günther,
  Mengzhou Xia, Weijia Shi, Xing~Han Lù, Jordan Clive, Gayatri Krishnakumar,
  Anna Maksimova, Silvan Wehrli, Maria Tikhonova, Henil Panchal, Aleksandr
  Abramov, Malte Ostendorff, Zheng Liu, Simon Clematide, Lester~James Miranda,
  Alena Fenogenova, Guangyu Song, Ruqiya~Bin Safi, Wen-Ding Li, Alessia
  Borghini, Federico Cassano, Hongjin Su, Jimmy Lin, Howard Yen, Lasse Hansen,
  Sara Hooker, Chenghao Xiao, Vaibhav Adlakha, Orion Weller, Siva Reddy, and
  Niklas Muennighoff.
\newblock Mmteb: Massive multilingual text embedding benchmark.
\newblock In \emph{International Conference on Learning Representations}.
  arXiv, 2025.
\newblock \doi{10.48550/arXiv.2502.13595}.
\newblock URL \url{https://arxiv.org/abs/2502.13595}.

\bibitem[Faysse et~al.(2025)Faysse, Sibille, Wu, Omrani, Viaud, Hudelot, and
  Colombo]{faysse2025colpali}
Manuel Faysse, Hugues Sibille, Tony Wu, Bilel Omrani, Gautier Viaud, C{\'e}line
  Hudelot, and Pierre Colombo.
\newblock {ColPali}: Efficient document retrieval with vision language models.
\newblock In \emph{The Thirteenth International Conference on Learning
  Representations}, 2025.
\newblock arXiv:2407.01449.

\bibitem[Georgiou(2026{\natexlab{a}})]{vultron2026}
Athos Georgiou.
\newblock {VultronRetrieverFlash-Qwen3.5-0.8B}: Small-tier late-interaction
  visual document retrieval at 320 dimensions.
\newblock
  \url{https://huggingface.co/vultr/VultronRetrieverFlash-Qwen3.5-0.8B},
  2026{\natexlab{a}}.

\bibitem[Georgiou(2026{\natexlab{b}})]{vultroncore}
Athos Georgiou.
\newblock {VultronRetrieverCore-Qwen3.5-4.5B}: Mid-tier late-interaction visual
  document retrieval at 320 dimensions.
\newblock \url{https://huggingface.co/vultr/VultronRetrieverCore-Qwen3.5-4.5B},
  2026{\natexlab{b}}.

\bibitem[{Granite Vision Team}(2025)]{granitevision2025}
{Granite Vision Team}.
\newblock Granite vision: a lightweight, open-source multimodal model for
  enterprise intelligence, 2025.

\bibitem[{Granite Vision Team} et~al.(2025){Granite Vision Team}, Karlinsky,
  Arbelle, Daniels, Nassar, Alfassi, Wu, Schwartz, Joshi, Kondic,
  et~al.]{team2025granite}
{Granite Vision Team}, Leonid Karlinsky, Assaf Arbelle, Abraham Daniels, Ahmed
  Nassar, Amit Alfassi, Bo~Wu, Eli Schwartz, Dhiraj Joshi, Jovana Kondic,
  et~al.
\newblock Granite vision: a lightweight, open-source multimodal model for
  enterprise intelligence.
\newblock \emph{arXiv preprint arXiv:2502.09927}, 2025.

\bibitem[G{\"u}nther et~al.(2025)G{\"u}nther, Sturua, Akram, Mohr, Ungureanu,
  Wang, Eslami, Martens, Werk, Wang, and Xiao]{gunther2025jinav4}
Michael G{\"u}nther, Saba Sturua, Mohammad~Kalim Akram, Isabelle Mohr, Andrei
  Ungureanu, Bo~Wang, Sedigheh Eslami, Scott Martens, Maximilian Werk, Nan
  Wang, and Han Xiao.
\newblock jina-embeddings-v4: Universal embeddings for multimodal multilingual
  retrieval.
\newblock In \emph{Proceedings of the 5th Workshop on Multilingual
  Representation Learning (MRL 2025)}, pp.\  531--550. Association for
  Computational Linguistics, November 2025.
\newblock \doi{10.18653/v1/2025.mrl-main.36}.
\newblock URL \url{https://aclanthology.org/2025.mrl-main.36/}.

\bibitem[H{\"o}nicke et~al.(2026)H{\"o}nicke, G{\"u}nther, Koukounas, Akram,
  Sturua, and Xiao]{jinav5omni2026}
Florian H{\"o}nicke, Michael G{\"u}nther, Andreas Koukounas, Mohammad~Kalim
  Akram, Saba Sturua, and Han Xiao.
\newblock jina-embeddings-v5-omni: Geometry-preserving embeddings via locked
  aligned towers.
\newblock \url{https://huggingface.co/jinaai/jina-embeddings-v5-omni-small},
  2026.
\newblock URL \url{https://arxiv.org/abs/2605.08384}.

\bibitem[Hu et~al.(2022)Hu, Shen, Wallis, Allen-Zhu, Li, Wang, Wang, and
  Chen]{hu2022lora}
Edward~J. Hu, Yelong Shen, Phillip Wallis, Zeyuan Allen-Zhu, Yuanzhi Li, Shean
  Wang, Lu~Wang, and Weizhu Chen.
\newblock {LoRA}: Low-rank adaptation of large language models.
\newblock In \emph{International Conference on Learning Representations}, 2022.

\bibitem[{IBM Granite Vision Team}(2026)]{granitevision5}
{IBM Granite Vision Team}.
\newblock granite-vision-5.0-4b.
\newblock \url{https://huggingface.co/ibm-granite/granite-vision-5.0-4b}, 2026.

\bibitem[Ilharco et~al.(2023)Ilharco, Ribeiro, Wortsman, Gururangan, Schmidt,
  Hajishirzi, and Farhadi]{ilharco2023taskarithmetic}
Gabriel Ilharco, Marco~Tulio Ribeiro, Mitchell Wortsman, Suchin Gururangan,
  Ludwig Schmidt, Hannaneh Hajishirzi, and Ali Farhadi.
\newblock Editing models with task arithmetic.
\newblock In \emph{The Eleventh International Conference on Learning
  Representations}, 2023.

\bibitem[Khattab \& Zaharia(2020)Khattab and Zaharia]{khattab2020colbert}
Omar Khattab and Matei Zaharia.
\newblock {ColBERT}: Efficient and effective passage search via contextualized
  late interaction over {BERT}.
\newblock In \emph{Proceedings of the 43rd International ACM SIGIR Conference
  on Research and Development in Information Retrieval}, 2020.

\bibitem[Kolavi \& Jain(2025)Kolavi and Jain]{m3dr2025}
Adithya~S Kolavi and Vyoman Jain.
\newblock {M3DR}: Towards universal multilingual multimodal document retrieval,
  2025.

\bibitem[Kondic et~al.(2026)Kondic, Li, Joshi, Sanchez, Wiesel, Abedin,
  Alfassy, Schwartz, et~al.]{chartnet2026}
Jovana Kondic, Pengyuan Li, Dhiraj Joshi, Isaac Sanchez, Ben Wiesel, Shafiq
  Abedin, Amit Alfassy, Eli Schwartz, et~al.
\newblock {ChartNet}: A million-scale, high-quality multimodal dataset for
  robust chart understanding.
\newblock In \emph{Proceedings of the IEEE/CVF Conference on Computer Vision
  and Pattern Recognition (CVPR)}, 2026.

\bibitem[Krimmel et~al.(2026)Krimmel, Milliken, Glazkova, Schall, Chaffin,
  Eslami, Golde, Bykov, Beamer, Yang, and Wang]{pplxembedv2}
Markus Krimmel, Louis Milliken, Ekaterina Glazkova, Maximilian Schall, Antoine
  Chaffin, Sedigheh Eslami, Jonas Golde, Denis Bykov, Claire Beamer, Jeremy
  Yang, and Bo~Wang.
\newblock {pplx-embed-v2-late} technical report.
\newblock \url{https://huggingface.co/perplexity-ai/pplx-embed-v2-late-0.6b},
  2026.
\newblock URL \url{https://arxiv.org/abs/2610.29523}.

\bibitem[Kusupati et~al.(2022)Kusupati, Bhatt, Rege, Wallingford, Sinha,
  Ramanujan, Howard-Snyder, Chen, Kakade, Jain, and
  Farhadi]{kusupati2024matryoshkarepresentationlearning}
Aditya Kusupati, Gantavya Bhatt, Aniket Rege, Matthew Wallingford, Aditya
  Sinha, Vivek Ramanujan, William Howard-Snyder, Kaifeng Chen, Sham Kakade,
  Prateek Jain, and Ali Farhadi.
\newblock Matryoshka representation learning.
\newblock In \emph{Advances in Neural Information Processing Systems},
  volume~35, 2022.
\newblock URL \url{https://arxiv.org/abs/2205.13147}.

\bibitem[Kwiatkowski et~al.(2019)Kwiatkowski, Palomaki, Redfield, Collins,
  Parikh, Alberti, Epstein, Polosukhin, Devlin, Lee, Toutanova, Jones, Kelcey,
  Chang, Dai, Uszkoreit, Le, and
  Petrov]{kwiatkowski-etal-2019-natural-questions}
Tom Kwiatkowski, Jennimaria Palomaki, Olivia Redfield, Michael Collins, Ankur
  Parikh, Chris Alberti, Danielle Epstein, Illia Polosukhin, Jacob Devlin,
  Kenton Lee, Kristina Toutanova, Llion Jones, Matthew Kelcey, Ming-Wei Chang,
  Andrew~M. Dai, Jakob Uszkoreit, Quoc Le, and Slav Petrov.
\newblock {Natural Questions: A Benchmark for Question Answering Research}.
\newblock \emph{Transactions of the ACL}, 2019.

\bibitem[Kydl{\'\i}{\v{c}}ek et~al.(2025)Kydl{\'\i}{\v{c}}ek, Penedo, and von
  Werra]{kydlicek2025finepdfs}
Hynek Kydl{\'\i}{\v{c}}ek, Guilherme Penedo, and Leandro von Werra.
\newblock {FinePDFs}.
\newblock \url{https://huggingface.co/datasets/HuggingFaceFW/finepdfs}, 2025.

\bibitem[Lauren{\c{c}}on et~al.(2024{\natexlab{a}})Lauren{\c{c}}on, Marafioti,
  Sanh, and Tronchon]{laurencon2024idefics3}
Hugo Lauren{\c{c}}on, Andr{\'e}s Marafioti, Victor Sanh, and L{\'e}o Tronchon.
\newblock Building and better understanding vision-language models: insights
  and future directions, 2024{\natexlab{a}}.

\bibitem[Lauren{\c{c}}on et~al.(2024{\natexlab{b}})Lauren{\c{c}}on, Tronchon,
  Cord, and Sanh]{laurencon2024idefics2}
Hugo Lauren{\c{c}}on, L{\'e}o Tronchon, Matthieu Cord, and Victor Sanh.
\newblock What matters when building vision-language models?,
  2024{\natexlab{b}}.

\bibitem[Li et~al.(2023)Li, Li, Savarese, and Hoi]{li2023blip2}
Junnan Li, Dongxu Li, Silvio Savarese, and Steven Hoi.
\newblock {BLIP-2}: Bootstrapping language-image pre-training with frozen image
  encoders and large language models.
\newblock In \emph{Proceedings of the 40th International Conference on Machine
  Learning}, 2023.

\bibitem[Li et~al.(2026)Li, Zhang, Long, Chen, Song, Bai, Yang, Xie, Yang, Liu,
  Zhou, and Lin]{qwen3vlembedding}
Mingxin Li, Yanzhao Zhang, Dingkun Long, Keqin Chen, Sibo Song, Shuai Bai,
  Zhibo Yang, Pengjun Xie, An~Yang, Dayiheng Liu, Jingren Zhou, and Junyang
  Lin.
\newblock {Qwen3-VL-Embedding} and {Qwen3-VL-Reranker}: A unified framework for
  state-of-the-art multimodal retrieval and ranking.
\newblock \url{https://huggingface.co/Qwen/Qwen3-VL-Embedding-2B},
  \url{https://huggingface.co/Qwen/Qwen3-VL-Embedding-8B}, 2026.
\newblock URL \url{https://arxiv.org/abs/2601.04720}.

\bibitem[Liu et~al.(2024)Liu, Wang, Yin, Molchanov, Wang, Cheng, and
  Chen]{liu2024dora}
Shih-Yang Liu, Chien-Yi Wang, Hongxu Yin, Pavlo Molchanov, Yu-Chiang~Frank
  Wang, Kwang-Ting Cheng, and Min-Hung Chen.
\newblock {DoRA}: Weight-decomposed low-rank adaptation.
\newblock In \emph{Proceedings of the 41st International Conference on Machine
  Learning}, volume 235 of \emph{Proceedings of Machine Learning Research},
  pp.\  32100--32121. PMLR, 2024.
\newblock URL \url{https://proceedings.mlr.press/v235/liu24bn.html}.

\bibitem[Loison et~al.(2026)Loison, Mac{\'e}, Edy, Xing, Balough, Moreira, Liu,
  Faysse, Hudelot, and Viaud]{vidorev3}
Ant{\'o}nio Loison, Quentin Mac{\'e}, Antoine Edy, Victor Xing, Tom Balough,
  Gabriel de Souza~P. Moreira, Bo~Liu, Manuel Faysse, Celine Hudelot, and
  Gautier Viaud.
\newblock {ViDoRe V3}: A comprehensive evaluation of retrieval augmented
  generation in complex real-world scenarios.
\newblock In \emph{Proceedings of the 64th Annual Meeting of the Association
  for Computational Linguistics (Volume 1: Long Papers)}, pp.\  16570--16600,
  San Diego, California, United States, July 2026. Association for
  Computational Linguistics.
\newblock \doi{10.18653/v1/2026.acl-long.755}.
\newblock URL \url{https://aclanthology.org/2026.acl-long.755/}.

\bibitem[Lu et~al.(2023)Lu, Qiu, Chang, Wu, Zhu, Rajpurohit, Clark, and
  Kalyan]{lu2023tabmwp}
Pan Lu, Liang Qiu, Kai-Wei Chang, Yingnian Wu, Song-Chun Zhu, Tanmay
  Rajpurohit, Peter Clark, and Ashwin Kalyan.
\newblock Dynamic prompt learning via policy gradient for semi-structured
  mathematical reasoning.
\newblock In \emph{International Conference on Learning Representations}, 2023.
\newblock URL \url{https://openreview.net/forum?id=DHyHRBwJUTN}.

\bibitem[Masry et~al.(2022)Masry, Long, Tan, Joty, and Hoque]{masry2022chartqa}
Ahmed Masry, Do~Xuan Long, Jia~Qing Tan, Shafiq Joty, and Enamul Hoque.
\newblock {C}hart{QA}: A benchmark for question answering about charts with
  visual and logical reasoning.
\newblock In Smaranda Muresan, Preslav Nakov, and Aline Villavicencio (eds.),
  \emph{Findings of the {A}ssociation for {C}omputational {L}inguistics: {ACL}
  2022}, pp.\  2263--2279, Dublin, Ireland, May 2022. Association for
  Computational Linguistics.
\newblock \doi{10.18653/v1/2022.findings-acl.177}.
\newblock URL \url{https://aclanthology.org/2022.findings-acl.177/}.

\bibitem[Mathew et~al.(2021)Mathew, Karatzas, and Jawahar]{mathew2021docvqa}
Minesh Mathew, Dimosthenis Karatzas, and C.~V. Jawahar.
\newblock {DocVQA}: A dataset for {VQA} on document images.
\newblock In \emph{Proceedings of the IEEE/CVF Winter Conference on
  Applications of Computer Vision (WACV)}, pp.\  2200--2209, January 2021.

\bibitem[Mathew et~al.(2022)Mathew, Bagal, Tito, Karatzas, Valveny, and
  Jawahar]{mathew2022infographicvqa}
Minesh Mathew, Viraj Bagal, Rub{\`e}n Tito, Dimosthenis Karatzas, Ernest
  Valveny, and C.~V. Jawahar.
\newblock {InfographicVQA}.
\newblock In \emph{Proceedings of the IEEE/CVF Winter Conference on
  Applications of Computer Vision (WACV)}, pp.\  1697--1706, January 2022.

\bibitem[Meng et~al.(2026)Meng, Jiang, Liu, Su, Yang, Fu, Qin, Thirukovalluru,
  Zhang, Chen, Xu, Xiong, Zhou, Chen, and Yavuz]{visdocood}
Rui Meng, Ziyan Jiang, Ye~Liu, Mingyi Su, Xinyi Yang, Yuepeng Fu, Can Qin,
  Raghuveer Thirukovalluru, Xuan Zhang, Zeyuan Chen, Ran Xu, Caiming Xiong,
  Yingbo Zhou, Wenhu Chen, and Semih Yavuz.
\newblock {VLM2Vec-V2}: Advancing multimodal embedding for videos, images, and
  visual documents.
\newblock \emph{Transactions on Machine Learning Research}, 2026.
\newblock URL \url{https://openreview.net/forum?id=TpU38jbKIJ}.

\bibitem[Methani et~al.(2020)Methani, Ganguly, Khapra, and
  Kumar]{methani2020plotqa}
Nitesh Methani, Pritha Ganguly, Mitesh~M. Khapra, and Pratyush Kumar.
\newblock {PlotQA}: Reasoning over scientific plots.
\newblock In \emph{Proceedings of the IEEE/CVF Winter Conference on
  Applications of Computer Vision (WACV)}, pp.\  1527--1536, March 2020.

\bibitem[Muennighoff et~al.(2023)Muennighoff, Tazi, Magne, and
  Reimers]{muennighoff2022mteb}
Niklas Muennighoff, Nouamane Tazi, Lo{\"\i}c Magne, and Nils Reimers.
\newblock {MTEB}: Massive text embedding benchmark.
\newblock In \emph{Proceedings of the 17th Conference of the {E}uropean Chapter
  of the {A}ssociation for {C}omputational {L}inguistics}, pp.\  2014--2037,
  Dubrovnik, Croatia, May 2023. Association for Computational Linguistics.
\newblock \doi{10.18653/v1/2023.eacl-main.148}.
\newblock URL \url{https://aclanthology.org/2023.eacl-main.148/}.

\bibitem[{Nomic Team}(2025)]{colnomic2025}
{Nomic Team}.
\newblock {Nomic Embed Multimodal}: Interleaved text, image, and screenshots
  for visual document retrieval.
\newblock \url{https://huggingface.co/nomic-ai/colnomic-embed-multimodal-7b},
  2025.
\newblock URL \url{https://nomic.ai/blog/posts/nomic-embed-multimodal}.

\bibitem[{NVIDIA}(2026)]{nemotron35embed}
{NVIDIA}.
\newblock {NVIDIA Nemotron 3.5 Embed} ({Nemotron-3.5-Embed-1B-BF16} and
  {Nemotron-3.5-Embed-8B-BF16}).
\newblock \url{https://huggingface.co/nvidia/Nemotron-3.5-Embed-1B-BF16},
  \url{https://huggingface.co/nvidia/Nemotron-3.5-Embed-8B-BF16}, 2026.

\bibitem[Or et~al.(2025)Or, Jain, Vega-Myhre, Cai, Hernandez, Zhang, Guessous,
  Kuznetsov, Puhrsch, Saroufim, Rao, Tran, and
  Samard{\v{z}}i{\'c}]{torchao2025}
Andrew Or, Apurva Jain, Daniel Vega-Myhre, Jesse Cai, Charles~David Hernandez,
  Zhenrui Zhang, Driss Guessous, Vasiliy Kuznetsov, Christian Puhrsch, Mark
  Saroufim, Supriya Rao, Thien Tran, and Aleksandar Samard{\v{z}}i{\'c}.
\newblock {TorchAO}: {PyTorch}-native training-to-serving model optimization.
\newblock In \emph{Proceedings of the ICML 2025 Workshop on Championing
  Opensource Development in Machine Learning (CODEML '25)}, 2025.
\newblock URL \url{https://arxiv.org/abs/2507.16099}.

\bibitem[Santhanam et~al.(2022)Santhanam, Khattab, Potts, and
  Zaharia]{santhanam2022plaid}
Keshav Santhanam, Omar Khattab, Christopher Potts, and Matei Zaharia.
\newblock {PLAID}: An efficient engine for late interaction retrieval.
\newblock In \emph{Proceedings of the 31st ACM International Conference on
  Information \& Knowledge Management}, 2022.

\bibitem[Teiletche et~al.(2026)Teiletche, Mac{\'e}, Conti, Loison, Viaud,
  Colombo, and Faysse]{modernvbert2025}
Paul Teiletche, Quentin Mac{\'e}, Max Conti, Ant{\'o}nio Loison, Gautier Viaud,
  Pierre Colombo, and Manuel Faysse.
\newblock {ModernVBERT}: Towards smaller visual document retrievers.
\newblock In \emph{International Conference on Machine Learning (ICML)}, 2026.
\newblock URL \url{https://icml.cc/virtual/2026/poster/63761}.

\bibitem[{TopK}(2026)]{topkembed}
{TopK}.
\newblock {topk-embed-v1-small}.
\newblock \url{https://huggingface.co/topk-io/topk-embed-v1-small}, 2026.

\bibitem[Tschannen et~al.(2025)Tschannen, Gritsenko, Wang, Naeem,
  Alabdulmohsin, Parthasarathy, Evans, Beyer, Xia, Mustafa, H{\'e}naff,
  Harmsen, Steiner, and Zhai]{tschannen2025siglip2}
Michael Tschannen, Alexey Gritsenko, Xiao Wang, Muhammad~Ferjad Naeem, Ibrahim
  Alabdulmohsin, Nikhil Parthasarathy, Talfan Evans, Lucas Beyer, Ye~Xia, Basil
  Mustafa, Olivier H{\'e}naff, Jeremiah Harmsen, Andreas Steiner, and Xiaohua
  Zhai.
\newblock {SigLIP} 2: Multilingual vision-language encoders with improved
  semantic understanding, localization, and dense features, 2025.

\bibitem[Turski et~al.(2023)Turski, Stanis{\l}awek, Kaczmarek, Dyda, and
  Grali{\'n}ski]{turski2023ccpdf}
Micha{\l} Turski, Tomasz Stanis{\l}awek, Karol Kaczmarek, Pawe{\l} Dyda, and
  Filip Grali{\'n}ski.
\newblock Ccpdf: Building a high quality corpus for visually rich documents
  from web crawl data.
\newblock In \emph{International conference on document analysis and
  recognition}, pp.\  348--365. Springer, 2023.

\bibitem[Vera et~al.(2026)Vera, Dua, Andonov, Smoot, Russo, Llanos, Matveev,
  Yoo, Black, Huang, Aklilu, Grootendorst, Sanseviero, Amarnath, Ballantyne,
  Li, Salz, Alabdulmohsin, Philippov, Quan, Cameron, Yvinec, Martins, Reid,
  Panyam, Wang, Ying, Shanbhogue, Dabral, Zhang, Ji, Wu, Zou, Chen, Zhang,
  Chen, Soulanille, Mamillapalli, Kuo, Law, Kreileder, Zheng, Shah, Shieh,
  Consul, Montes, Hess, Sevilla, Samari, McCloskey, Potetz, Cer, Kathareios,
  {\'A}brego, Zhang, Ma, Wu, Mosley, Suganthan, Meng, Gao, Cao, Chen, Chen,
  Ding, Byrne, Brick, Lacombe, Yin, Warkentin, Joulin, Duerig, and
  Seyedhosseini]{embeddinggemma2}
Henrique~Schechter Vera, Sahil Dua, Jovan Andonov, Sara Smoot, Sebastian Russo,
  Ivan Llanos, Max Matveev, Jae~Hyeon Yoo, Kat Black, Shih-Cheng Huang, Josiah
  Aklilu, Maarten Grootendorst, Omar Sanseviero, Chandramouli Amarnath, Ian
  Ballantyne, Zhe Li, Daniel Salz, Ibrahim Alabdulmohsin, Ilya Philippov,
  Kristen Quan, Glenn Cameron, Edouard Yvinec, Luciano Martins, Doug Reid,
  Sindhu~Raghuram Panyam, Weiyi Wang, Hengxuan Ying, Madhuri Shanbhogue,
  Tanmaya Dabral, Shanfeng Zhang, Juyeong Ji, Renjie Wu, Joe Zou, Xu~Chen,
  Yunan Zhang, Yu-Hui Chen, Matthew Soulanille, Vinod Mamillapalli, Yi-Chun
  Kuo, Wai~Hon Law, Matt Kreileder, Alice Zheng, Suril Shah, Mogan Shieh, Pooja
  Consul, Alberto Montes, Andreas Hess, Andres Sevilla, Babak Samari, Brendan
  McCloskey, Brian Potetz, Daniel Cer, Georgios Kathareios,
  Gustavo~Hern{\'a}ndez {\'A}brego, Hesen Zhang, Ji~Ma, Jiaxing Wu, Keegan
  Mosley, Paul Suganthan, Rui Meng, Shen Gao, Shengcao Cao, Yi-Ting Chen,
  Yichang Chen, Zhongli Ding, Kendra Byrne, Cormac Brick, Olivier Lacombe, Qin
  Yin, Tris Warkentin, Armand Joulin, Tom Duerig, and Mojtaba Seyedhosseini.
\newblock {EmbeddingGemma 2}: Unified and efficient multimodal representations.
\newblock \url{https://huggingface.co/google/embeddinggemma-2}, 2026.
\newblock URL \url{https://arxiv.org/abs/2610.37364}.

\bibitem[{ViDoRe}(2024{\natexlab{a}})]{colqwen25}
{ViDoRe}.
\newblock {ColQwen2.5-v0.2}.
\newblock \url{https://huggingface.co/vidore/colqwen2.5-v0.2},
  2024{\natexlab{a}}.

\bibitem[{ViDoRe}(2024{\natexlab{b}})]{colsmol2025}
{ViDoRe}.
\newblock {ColSmol-500M}.
\newblock \url{https://huggingface.co/vidore/colSmol-500M}, 2024{\natexlab{b}}.

\bibitem[Wang et~al.(2026)Wang, Wen, Ji, Zhang, Qiao, and Sun]{evie}
Zifei Wang, Wei Wen, Qiang Ji, Qian-Wen Zhang, Ruizhi Qiao, and Xing Sun.
\newblock {EVIE}: Evidence-vector-informed embeddings for visual document
  retrieval.
\newblock \url{https://huggingface.co/tencent/EVIE-4.5B}, 2026.
\newblock URL \url{https://arxiv.org/abs/2610.11553}.

\bibitem[Warner et~al.(2025)Warner, Chaffin, Clavi{\'e}, Weller, Hallstr{\"o}m,
  Taghadouini, Gallagher, Biswas, Ladhak, Aarsen, Cooper, Adams, Howard, and
  Poli]{warner2024modernbert}
Benjamin Warner, Antoine Chaffin, Benjamin Clavi{\'e}, Orion Weller, Oskar
  Hallstr{\"o}m, Said Taghadouini, Alexis Gallagher, Raja Biswas, Faisal
  Ladhak, Tom Aarsen, Nathan Cooper, Griffin Adams, Jeremy Howard, and Iacopo
  Poli.
\newblock Smarter, better, faster, longer: A modern bidirectional encoder for
  fast, memory efficient, and long context finetuning and inference.
\newblock In \emph{Proceedings of the 63rd Annual Meeting of the Association
  for Computational Linguistics (Volume 1: Long Papers)}, Vienna, Austria,
  2025. Association for Computational Linguistics.

\bibitem[Wasserman et~al.(2025)Wasserman, Pony, Naparstek, Goldfarb, Schwartz,
  Barzelay, and Karlinsky]{realmmrag2025}
Navve Wasserman, Roi Pony, Oshri Naparstek, Adi~Raz Goldfarb, Eli Schwartz, Udi
  Barzelay, and Leonid Karlinsky.
\newblock {REAL}-{MM}-{RAG}: A real-world multi-modal retrieval benchmark.
\newblock In \emph{Proceedings of the 63rd Annual Meeting of the Association
  for Computational Linguistics (Volume 1: Long Papers)}, pp.\  31660--31683,
  Vienna, Austria, July 2025. Association for Computational Linguistics.
\newblock \doi{10.18653/v1/2025.acl-long.1528}.
\newblock URL \url{https://aclanthology.org/2025.acl-long.1528/}.

\bibitem[{webAI}(2026)]{webaicolvec}
{webAI}.
\newblock {webAI-ColVec1.1-4b}: A bidirectional multi-vector model for visual
  document retrieval.
\newblock \url{https://huggingface.co/webAI-Official/webAI-ColVec1.1-4b}, 2026.

\bibitem[Wortsman et~al.(2022)]{wortsman2022soups}
Mitchell Wortsman et~al.
\newblock Model soups: averaging weights of multiple fine-tuned models improves
  accuracy without increasing inference time.
\newblock In \emph{Proceedings of the 39th International Conference on Machine
  Learning}, 2022.

\bibitem[Xiao et~al.(2026)Xiao, Ma, Gu, Chen, Chen, Ordonez, and
  Mohan]{metaembed2025}
Zilin Xiao, Qi~Ma, Mengting Gu, Chun-cheng~Jason Chen, Xintao Chen, Vicente
  Ordonez, and Vijai Mohan.
\newblock {MetaEmbed}: Scaling multimodal retrieval at test-time with flexible
  late interaction.
\newblock In \emph{International Conference on Learning Representations
  (ICLR)}, 2026.
\newblock URL \url{https://openreview.net/forum?id=yKDqg9HwZX}.

\bibitem[Yao et~al.(2025)Yao, Huang, Wu, Zhang, Wang, Liu, Wang, Song, Feng,
  Shen, and Tao]{yao2024mulberry}
Huanjin Yao, Jiaxing Huang, Wenhao Wu, Jingyi Zhang, Yibo Wang, Shunyu Liu,
  Yingjie Wang, Yuxin Song, Haocheng Feng, Li~Shen, and Dacheng Tao.
\newblock Mulberry: Empowering {MLLM} with o1-like reasoning and reflection via
  collective {M}onte {C}arlo tree search.
\newblock In \emph{Advances in Neural Information Processing Systems}, 2025.

\bibitem[Yu et~al.(2025)Yu, Tang, Xu, Cui, Ran, Yan, Liu, Wang, Han, Liu, and
  Sun]{visrag}
Shi Yu, Chaoyue Tang, Bokai Xu, Junbo Cui, Junhao Ran, Yukun Yan, Zhenghao Liu,
  Shuo Wang, Xu~Han, Zhiyuan Liu, and Maosong Sun.
\newblock {VisRAG}: Vision-based retrieval-augmented generation on
  multi-modality documents.
\newblock In \emph{International Conference on Learning Representations
  (ICLR)}, pp.\  21074--21098, 2025.
\newblock URL
  \url{https://proceedings.iclr.cc/paper_files/paper/2025/hash/3640a1997a4c9571cea9db2c82e1fc35-Abstract-Conference.html}.

\bibitem[Zhang et~al.(2023)Zhang, Thakur, Ogundepo, Kamalloo, Alfonso-Hermelo,
  Li, Liu, Rezagholizadeh, and Lin]{zhang-etal-2023-miracl}
Xinyu Zhang, Nandan Thakur, Odunayo Ogundepo, Ehsan Kamalloo, David
  Alfonso-Hermelo, Xiaoguang Li, Qun Liu, Mehdi Rezagholizadeh, and Jimmy Lin.
\newblock {MIRACL}: A multilingual retrieval dataset covering 18 diverse
  languages.
\newblock \emph{Transactions of the Association for Computational Linguistics},
  11:\penalty0 1114--1131, 2023.
\newblock \doi{10.1162/tacl_a_00595}.
\newblock URL \url{https://aclanthology.org/2023.tacl-1.63}.

\bibitem[Zheng et~al.(2021)Zheng, Burdick, Popa, Zhong, and Wang]{fintabnet}
Xinyi Zheng, Douglas Burdick, Lucian Popa, Xu~Zhong, and Nancy Xin~Ru Wang.
\newblock Global table extractor {(GTE):} {A} framework for joint table
  identification and cell structure recognition using visual context.
\newblock In \emph{{IEEE} Winter Conference on Applications of Computer Vision,
  {WACV} 2021, Waikoloa, HI, USA, January 3-8, 2021}, pp.\  697--706. {IEEE},
  2021.
\newblock \doi{10.1109/WACV48630.2021.00074}.
\newblock URL \url{https://doi.org/10.1109/WACV48630.2021.00074}.

\end{thebibliography}
\bibliographystyle{iclr2025_conference}

\newpage
\appendix

\section{Contributions}
\label{sec:contributions}
\modelfullname{} is the work of the Granite Embedding team at IBM Research, with Parul Awasthy and Radu Florian as the challenge leads on the project overall. We are very grateful to the Granite Vision team: the vision tower of \modelfullname{} starts from the weights of granite-vision-5.0-4b. We look forward to even better models!

\newcommand{\contribrole}[2]{\par\smallskip\noindent{\small\textsc{#1}}\par\noindent #2\par}
\contribrole{Modality Alignment}{Yahan Yang}
\contribrole{Retriever training}{Parul Awasthy, Ken Barker, Aashka Trivedi, Yahan Yang, Yushu Yang}
\contribrole{Data and evaluation}{Parul Awasthy, Radu Florian, Martin Franz, Bhavani Iyer, Yulong Li, Aashka Trivedi, Todd Ward, Yahan Yang, Yushu Yang}
\contribrole{Product management}{Madison Lee}
\contribrole{Model grooming, shoeing and speed training}{Radu Florian}
\contribrole{Technical leadership}{Parul Awasthy, Radu Florian, Luis Lastras}
\section{Training Data}
\label{app:data}

\begin{table}[ht]
  \centering
  \small
  {\renewcommand{\arraystretch}{1.15}
  \begin{tabular}{llr}
    \toprule
    Step & Dataset & Rows \\
    \midrule
    \multirow{6}{*}{1--2}
      & The Cauldron (35 subsets)  & 1{,}591{,}488 \\
      & DocFM QA (multilingual)    & 1{,}358{,}721 \\
      & Docmatix                   & 1{,}273{,}215 \\
      & CLEVR-Math                &   556{,}082 \\
      & Molmo2-SynMultiImageQA     &   191{,}823 \\
      & OK-VQA                     &     9{,}009 \\
    \midrule
    \multirow{2}{*}{3}
      & FinePDFs QA                &   348{,}368 \\
      & FinePDFs long descriptions &   930{,}106 \\
    \midrule
    \multicolumn{2}{l}{\textbf{Total}} & \textbf{6{,}258{,}812} \\
    \bottomrule
  \end{tabular}}
  \caption{Modality-alignment data. Steps 1 and 2 use the same data; step 3 uses FinePDFs only, in 22 languages. Samples with more than 2 images in the datasets are removed.}
  \label{tab:alignment-data}
\end{table}

We list the data used for modality alignment in Table~\ref{tab:alignment-data}, and multimodal distillation in Table~\ref{tab:distill-corpora}. For text, we use the non-synthetic subset of triplet data described in Section 2.3 of \cite{awasthy2026graniteembeddingmultilingualr2}.

\begin{table}[t]
  \centering
  \footnotesize
  {\setlength{\tabcolsep}{4pt}
  \begin{tabular}{lrr>{\raggedright\arraybackslash}p{4.6cm}}
    \toprule
    Corpus & Documents & Training rows & Notes \\
    \midrule
    DocFM\citep{team2025granite} & 1{,}010{,}212 & 1{,}198{,}958 & Gemma-4-26B-it-mined hard negatives \\
    ChartNet \citep{chartnet2026} & 351{,}189 & 706{,}302 & English and translations into six languages \\
    Nemotron subsets & 964{,}717 & 1{,}866{,}837 & Queries from the corpus \\
    \quad Wikimedia captions (10 languages) & 500{,}000 & 1{,}399{,}948 & 50k images per language \\
    \quad Mulberry \citep{yao2024mulberry} (CoT) & 295{,}095 & 295{,}095 & Cleaned chain-of-thought queries \\
    \quad ChartQA \citep{masry2022chartqa} (CoT) & 45{,}710 & 45{,}710 & Gemma-generated queries \\
    \quad DocVQA \citep{mathew2021docvqa} (CoT) & 36{,}333 & 36{,}333 & Cleaned chain-of-thought queries \\
    \quad TabMWP \citep{lu2023tabmwp} (CoT) & 20{,}305 & 20{,}305 & Cleaned chain-of-thought queries \\
    \quad InfographicsVQA \citep{mathew2022infographicvqa} (CoT) & 19{,}548 & 19{,}548 & Gemma-generated queries \\
    \quad MapQA \citep{chang2022mapqa} (CoT) & 16{,}764 & 16{,}764 & Cleaned chain-of-thought queries \\
    \quad PlotQA \citep{methani2020plotqa} (CoT) & 16{,}256 & 16{,}256 & Cleaned chain-of-thought queries \\
    \quad FinTabNet \citep{fintabnet} (CoT) & 8{,}352 & 8{,}352 & Cleaned chain-of-thought queries \\
    \quad Paper-judge CCPDF pages \citep{turski2023ccpdf} & 5{,}916 & 7{,}991 & In-house synthetic queries \\
    \quad Paper-judge CCPDF figures \citep{turski2023ccpdf}& 438 & 535 & In-house synthetic queries \\
    Other charts and tables & 97{,}041 & 97{,}041 &  \\
    \quad ChartCap & 48{,}835 & 48{,}835 & Chart captions \\
    \quad FinTab & 48{,}206 & 48{,}206 & Financial tables \\
    Other document retrieval & 153{,}648 & 180{,}431 &  \\
    \quad VisRAG \citep{visrag} & 75{,}008 & 75{,}008 &  \\
    \quad ColPali DocVQA \citep{faysse2025colpali} & 39{,}302 & 66{,}085 & Two hard-negative sets \\
    \quad HMP synthetic & 21{,}000 & 21{,}000 &  \\
    \quad Jina commons & 11{,}170 & 11{,}170 &  \\
    \quad JDocQA & 7{,}168 & 7{,}168 & Japanese \\
    Synthetic multilingual VQA & 197{,}618 & 1{,}182{,}719 & Gemma-4-31B-it queries; en, de, es, fr, it, pt \\
    \quad PLOS papers & 68{,}000 & 405{,}011 &  \\
    \quad Financial tables & 47{,}423 & 284{,}538 &  \\
    \quad Report contents & 30{,}186 & 181{,}116 &  \\
    \quad AAAI papers & 26{,}673 & 160{,}038 &  \\
    \quad CVPR papers & 15{,}336 & 92{,}016 &  \\
    \quad Code images & 10{,}000 & 60{,}000 &  \\
    FinePDFs \citep{kydlicek2025finepdfs} pages (22 languages) & 848{,}344 & 1{,}696{,}688 & Long descriptions and generated queries \\
    Multi-page PDFs \citep{turski2023ccpdf} & 124{,}722 & 220{,}750 & In-house synthetic queries \\
    \quad Clustered, per language (6 languages) & 108{,}863 & 108{,}863 &  \\
    \quad Combined (14 query sets) & 15{,}859 & 111{,}887 & Translated and per-domain queries \\
    Product exports & 153{,}018 & 109{,}814 &  \\
    \quad Cross-lingual (v2) & 87{,}269 & 42{,}819 &  \\
    \quad Per language (18 languages) & 65{,}749 & 66{,}995 &  \\
    MIRACL \citep{zhang-etal-2023-miracl} rendered (16 languages) & 60{,}370 & 25{,}333 & Passages rendered as images \\
    FinePDFs papers \citep{kydlicek2025finepdfs}& 19{,}204 & 22{,}750 & In-house synthetic queries \\
    \quad CVPR & 8{,}338 & 9{,}432 &  \\
    \quad PLOS & 6{,}965 & 8{,}271 &  \\
    \quad AAAI & 3{,}901 & 5{,}047 &  \\
    English and French PDFs \citep{kydlicek2025finepdfs} & 9{,}344 & 10{,}856 & In-house synthetic queries \\
    \quad English & 5{,}114 & 5{,}937 &  \\
    \quad French & 4{,}230 & 4{,}919 &  \\
    \midrule
    \textbf{Total} & \textbf{3{,}989{,}427} & \textbf{7{,}318{,}479} & \\
    \bottomrule
  \end{tabular}}
  \caption{Multimodal distillation corpora. \emph{Documents}: distinct positive pages or images used in training, translations of the same pages counted once. \emph{Training rows}: examples built from each corpus's hard-negative file (lower than documents when an example has several positives).}
  \label{tab:distill-corpora}
\end{table}

\section{Training Hyperparameters}
\label{app:hyperparameters}

We show the Modality Alignment Hyperparameters in Table~\ref{tab:alignment-hparams} and the Distillation Hyperparameters in Table~\ref{tab:hyperparameters}. We also discuss the merge weight $\alpha$ toward \textmodelname{} (Section~\ref{sec:merge}) in Table~\ref{tab:merge-alpha}.

\begin{table}[h]
    \centering
    \begin{minipage}[t]{0.47\textwidth}
    \vspace{0pt}
    \centering
    {\renewcommand{\arraystretch}{1.3}
    \begin{adjustbox}{max width=\linewidth}
    \begin{tabular}{l|cc}
    \toprule
            & Text warmup & Multimodal distillation \\
    \midrule
        Learning rate & $2\times10^{-5}$ & $8\times10^{-5}$ \\
        Warmup ratio & 0.05 & 0.05\\
        Batch size (per GPU $\times$ GPUs) & $8 \times 8 = 64$ & $8 \times 8 = 64$ \\
        Max Steps  & 10{,}000  & 258{,}159 \\
        Max.\ sequence length & 1{,}024 & 2{,}048 \\
        Passages per query & \multicolumn{2}{c}{2 (positive + 1 hard negative)} \\
        Distillation temperature $\tau$ & 0.02 & 0.02 \\
        Dataset sampling temperature & 2 & 2 \\
        LoRA rank / alpha / dropout & 32 / 32 / 0.1 & 32 / 32 / 0.1 \\
        Precision & bf16 & bf16 \\
    \bottomrule
    \end{tabular}
    \end{adjustbox}}
    \caption{Distillation hyperparameters.}
    \label{tab:hyperparameters}
    \end{minipage}\hfill
    \begin{minipage}[t]{0.50\textwidth}
    \vspace{0pt}
    \centering
    {\renewcommand{\arraystretch}{1.3}
    \begin{adjustbox}{max width=\linewidth}
    \begin{tabular}{lccc}
    \toprule
    & Step 1 & Step 2 & Step 3 \\
    \midrule
    Initialization            & fresh              & Step 1                & Step 2 \\
    Training steps            & 5{,}500            & 2{,}500               & 2{,}500 \\
    MLM probability           & 0.5                & 0.5                   & 0.5 \\
    Max / min image patches   & 2{,}880 / 576      & 5{,}760 / 576         & 2{,}880 / 576 \\
    Max.\ sequence length     & 5{,}500            & 5{,}500               & 5{,}500 \\
    Trainable modules         & \multicolumn{2}{c}{text DoRA, connector} & + vision DoRA, patch emb. \\
    DoRA rank / $\alpha$      & 32 / 32            & 32 / 32               & 32 / 32 \\
    Peak learning rate        & $1\times10^{-4}$   & $1\times10^{-4}$      & $5\times10^{-5}$ \\
    Warmup / decay steps      & 275 / 1{,}100      & 275 / 1{,}100         & 100 / 1{,}100 \\
    Global batch size         & 2{,}048            & 2{,}048               & 2{,}048 \\
    \bottomrule
    \end{tabular}
    \end{adjustbox}}
    \caption{Modality-alignment hyperparameters. All steps use AdamW ($\beta_1{=}0.9$, $\beta_2{=}0.999$), weight decay 0.1, gradient clipping 1.0, bf16, and a warmup--stable--decay schedule with $1-\sqrt{\cdot}$ decay.}
    \label{tab:alignment-hparams}
    \end{minipage}
\end{table}

\begin{table}[h]
    \centering
    \small
    \begin{tabular}{l|c|c|cc}
    \toprule
        & ViDoRe v1 & & \multicolumn{2}{c}{$\Delta$ NQ nDCG@10} \\
        $\alpha$ & MV nDCG@5 & $\Delta$ & single vector & multi-vector \\
        \hline
        0 & 0.830 & -- & -- & -- \\
        \textbf{0.1} & \textbf{0.824} & \textbf{$-$0.006} & \textbf{+0.017} & \textbf{+0.014} \\
        0.2 & 0.807 & $-$0.023 & +0.031 & +0.017 \\
        0.3 & 0.782 & $-$0.048 & +0.040 & +0.020 \\
        0.5 & 0.718 & $-$0.112 & +0.051 & +0.023 \\
        1.0 & 0.503 & $-$0.327 & +0.057 & +0.022 \\
    \bottomrule
    \end{tabular}
    \caption{Choosing the merge weight $\alpha$ toward \textmodelname{}, on an earlier checkpoint. Images: multi-vector ViDoRe v1 (10 tasks). Text: change in nDCG@10 on a reduced Natural Questions dev-test pool (225,669 passages, 2,578 queries). $\alpha=1$ is the text encoder of \textmodelname{} with the student's multi-vector projection.}
    \label{tab:merge-alpha}
\end{table}

\section{Speed and Accuracy of All Models}
\label{app:speed}

Table~\ref{tab:speed} lists the throughput and scores behind Figure~\ref{fig:pareto}.

\begin{table}[t]
    \centering
    \footnotesize
    {\renewcommand{\arraystretch}{1.15}
    \begin{adjustbox}{max width=\textwidth}
    \begin{tabular}{l|r|c|r|c|c}
    \toprule
        Model & Params & Mode & Pages/s $\uparrow$ & ViDoRe v1--v3 & Avg.\ (8) \\
        \hline
        ColUltraVBERT & 0.26B & MV & 6.7 & 0.631 & 0.658 \\
        ColSmol-500M & 0.50B & MV & 13.5 & 0.538 & 0.581 \\
        pplx-embed-v2-late-0.6b & 0.59B & MV & 17.4 & 0.721 & 0.753 \\
        EmbeddingGemma-2$^\ddagger$ & 0.74B & SV & 28.2 & 0.565 & 0.606 \\
        Vultron Flash & 0.85B & MV & 17.1 & 0.684 & 0.710 \\
        Nemotron-3.5-Embed-1B & 1.55B & SV & 2.7 & 0.682 & 0.720 \\
        jina-embeddings-v5-omni-small & 1.63B & SV & 10.7 & 0.589 & 0.623 \\
        Qwen3-VL-Embedding-2B & 2.13B & SV & 13.1 & 0.671 & 0.664 \\
        topk-embed-v1-small & 2.22B & MV & 11.6 & 0.754 & 0.770 \\
        Argus-ColQwen3.5-2B & 2.32B & MV & 9.4 & 0.711 & 0.721 \\
        ColQwen2.5-v0.2 & 3.00B & MV & 9.8 & 0.685 & 0.707 \\
        jina-embeddings-v4$^\dagger$ & 3.93B & SV & 4.4 & 0.614 & 0.660 \\
        EVIE-4.5B & 4.54B & MV & 10.5 & \textbf{0.772} & \textbf{0.799} \\
        webAI-ColVec1.1-4b & 4.54B & MV & 7.6 & 0.739 & 0.777 \\
        VultronRetriever-Core-4.5B & 4.54B & MV & 9.4 & 0.743 & 0.776 \\
        Nemotron-ColEmbed-VL-4B-v2 & 4.80B & MV & 12.2 & 0.725 & 0.753 \\
        ColNomic-Embed-Multimodal-7B & 7.00B & MV & 8.9 & 0.704 & 0.751 \\
        Qwen3-VL-Embedding-8B & 8.14B & SV & 8.2 & 0.714 & 0.734 \\
        Nemotron-3.5-Embed-8B & 8.38B & SV & 2.6 & 0.713 & 0.751 \\
        Nemotron-ColEmbed-VL-8B-v2 & 8.70B & MV & 10.3 & 0.737 & 0.778 \\
        \hline
        \modelname{} & 0.75B & MV & \textbf{35.3} & 0.684 & 0.721 \\
        \modelname{} & 0.75B & SV & \textbf{35.3} & 0.577 & 0.601 \\
        \textcolor{gray}{Teacher (Granite Vision 4.1)} & \textcolor{gray}{4.00B} & \textcolor{gray}{MV} & \textcolor{gray}{13.1} & \textcolor{gray}{0.741} & \textcolor{gray}{0.769} \\
    \bottomrule
    \end{tabular}
    \end{adjustbox}}
    \caption{Data behind Figure~\ref{fig:pareto}. Throughput is page images encoded per second on one NVIDIA H200 in bf16, measured with the mteb harness for every model on ViDoRe v1 and v2 (14 tasks). Only \modelname{} uses the faster inference path of Section~\ref{subsec:inference-speedup}; every other model, including the teacher, which shares our image-processing pipeline, runs with its released preprocessing. MV: multi-vector (MaxSim); SV: single vector (cosine). $^\dagger$Images capped at 602{,}112 pixels. $^\ddagger$With the document prompt from its paper and 280 soft tokens per page, timed under transformers 5.19. Under transformers 5.21 it encodes 21.3 pages/s on the same node, because CPU image preprocessing is slower in transformers 5.21 and later; the outputs are identical.}
    \label{tab:speed}
\end{table}

\section{Comparison with EmbeddingGemma-2}
\label{app:vs-gemma}

EmbeddingGemma-2 is a strong model: it embeds text, images and audio in one space, and its single vector averages 0.606 over the eight benchmarks, slightly ahead of our single-vector head (0.601), while using far fewer parameters per page. The two models have almost the same number of parameters (748.3M and 744.4M) but spend them differently (Table~\ref{tab:vs-gemma}). EmbeddingGemma-2 has a smaller vision tower (168M against 417M) and pools each image to 280 tokens, while \modelname{} reads up to 2{,}880 patches at native aspect ratio and passes 144 to 720 visual tokens to its text encoder. Its 306M-parameter audio encoder and both models' vocabulary tables are idle when encoding a page image, so about 305M parameters are active per page for EmbeddingGemma-2 against about 547M for \modelname. The extra capacity and visual tokens are what the multi-vector head turns into a 0.115 higher average. Despite its larger active size, \modelname{} encodes 25\% more pages per second (Table~\ref{tab:speed}); Section~\ref{subsec:inference-speedup} shows that this lead comes from its inference path rather than its network.

\begin{table}[h]
    \centering
    \small
    {\renewcommand{\arraystretch}{1.25}
    \begin{adjustbox}{max width=\textwidth}
    \begin{tabular}{l|l|l}
    \toprule
        & \modelname & EmbeddingGemma-2 \\
        \hline
        Vision tower & 417.1M, SigLIP2 so400m NaFlex, 27 layers & 167.7M, Gemma ViT, 16 layers \\
        Image input & 576--2{,}880 patches of 16\,px (up to 0.74\,MP), aspect ratio kept & up to $\sim$0.65\,MP, aspect ratio kept \\
        Visual tokens & 144--720 per page (window Q-Former, 19.4M) & 280 per image ($3\times3$ pooling) \\
        Text transformer & 110.3M, ModernBERT, 22 layers & 136.8M, Gemma, 24 layers \\
        Vocabulary & 201.4M ($262{,}192\times768$) & 134.2M ($262{,}144\times512$) \\
        Audio encoder & -- & 305.7M \\
        \hline
        Total parameters & 748.3M & 744.4M \\
        Active per page encode & $\sim$547M & $\sim$305M \\
        Pages/s (H200) & 35.3 & 28.2 \\
        ViDoRe v1--v3 avg. & 0.684 (multi-vector), 0.577 (single) & 0.565 \\
        Avg.\ over 8 benchmarks & 0.721 (multi-vector), 0.601 (single) & 0.606 \\
    \bottomrule
    \end{tabular}
    \end{adjustbox}}
    \caption{\modelname{} and EmbeddingGemma-2 compared. Parameter counts are read from the checkpoints; throughput and scores are from Table~\ref{tab:speed}.}
    \label{tab:vs-gemma}
\end{table}

\section{Speeding Up Encoding Outside the Model}
\label{app:speedup-extra}

Section~\ref{subsec:inference-speedup} describes processing changes applied to \modelname{} to improve throughput.
We further test how far page encoding could be pushed without touching the weights or architecture of the model on an earlier tiled checkpoint (up to ten 384-pixel tiles per page), applying the same changes to EmbeddingGemma-2 in Table~\ref{tab:speedup-extra}. Of the following changes discussed, only GPU processing is shipped with the released \modelname{}.

\begin{table}[h]
    \centering
    \small
    {\renewcommand{\arraystretch}{1.2}
    \begin{tabular}{l|rr|cc}
    \toprule
        & \multicolumn{2}{c|}{Pages/s $\uparrow$} & \multicolumn{2}{c}{$\Delta$ nDCG@5, dev set} \\
        Configuration & Tiled checkpoint & EmbeddingGemma-2 & multi-vector & single \\
        \midrule
        As released & 18.9 & 28.2 & -- & -- \\
        + GPU preprocessing & 26.8 & 40.1 & $+0.0007$ & $-0.0002$ \\
        + page decoding overlapped with encoding & 40.7 & 57.5 & \multicolumn{2}{c}{unchanged} \\
        + FP8 vision tower, compiled & \textbf{45.6} & \textbf{61.2} & $+0.0018$ & $-0.0006$ \\
        \midrule
        \multicolumn{5}{l}{\emph{Not adopted}} \\
        \quad + FP8 text tower, compiled & 47.2 & 62.0 & $-0.0002$ & $-0.0030$ \\
        \quad at most 6 tiles per page & 52.3 & -- & $-0.0038$ & $-0.0012$ \\
        \quad at most 4 tiles per page & 55.5 & -- & $-0.0185$ & $-0.0045$ \\
    \bottomrule
    \end{tabular}}
    \caption{Speed-ups outside the model, on the earlier tiled checkpoint and on EmbeddingGemma-2 with the same treatment. Each adopted row adds to the one above; each row under ``not adopted'' adds to the FP8 vision tower. Pages/s on one H200 over ViDoRe v1 and v2. Quality changes are macro nDCG@5 differences against the previous configuration on our dev set, which contains no ViDoRe data.}
    \label{tab:speedup-extra}
\end{table}

\paragraph{Timing Profile.} We find that, on a single H200 GPU, the profile is limited by the CPU: per page, 20.9\,ms of image preprocessing and 6.4\,ms of other host-side work against 23.4\,ms of GPU compute. After moving preprocessing to the GPU, page decoding becomes the limiting factor: the evaluation pipeline decodes each page only when the next batch is requested, taking 27\,ms per DocVQA page.

\paragraph{Changes.}
\begin{itemize}
    \item \textbf{GPU preprocessing.} Pages are converted to 8-bit tensors on a thread pool, copied to the GPU through pinned memory one batch ahead, and resized and tiled there. This is the change the final model adopted, inside its image processor.
    \item \textbf{Overlapped decoding.} A background thread fetches and decodes the next batch while the current one encodes. This is the largest single gain (1.5$\times$ for both models), and requires changing only the data loader. However, we find that decoding remains the slowest stage on large pages even with this change.
    \item \textbf{FP8 vision tower.} The linear layers of the vision tower are quantized to FP8 after loading \citealp{torchao2025}, and the tower is recompiled with \texttt{torch}. 
\end{itemize}

\paragraph{What did not help.} INT8 in the place of FP8 brings no improvement on the H200. Quantizing the text tower to FP8 as well adds  3.5\% throughput at the cost of single vector performance. Fewer tiles per page trade resolution for speed, but also decreases performance for both heads.

\paragraph{Outcome.} Together, the three adopted changes make the tiled checkpoint 2.4$\times$ faster and EmbeddingGemma-2 2.2$\times$ faster, with every quality change within the 0.0025 points on the dev set. Further improvements could be brought by fewer visual tokens or a smaller vision tower.

\section{Text Compatibility with granite-embedding-311m-multilingual-r2}
\label{app:text-compat}

\modelname{} is built on the text encoder of \textmodelname{}, and several parts of its training keep the two models close. During modality alignment, a text-view distillation term penalizes any drift of the text \texttt{[CLS]} embedding away from that of the frozen \textmodelname{}, and from the second alignment step an image-view term pulls a page's \texttt{[CLS]} toward the \textmodelname{} embedding of its text (Section~\ref{sec:alignment}, Equation~\ref{eq:align-loss}). During distillation, the base text-encoder weights stay frozen and only rank-32 LoRA adapters are trained (Section~\ref{sec:kd}), and text data is distilled from the text teacher of Granite Embedding Multilingual R2. Finally, the released text encoder is merged back with the \textmodelname{} weights (Section~\ref{sec:merge}).

This raises a practical question: \textbf{are the single-vector embeddings of the two models interchangeable on text?} If they are, a deployment that already holds a \textmodelname{} text index can add page images encoded by \modelname{} without re-encoding its text.

\paragraph{Setup.} We compare the single-vector heads only (the 768-dimensional \texttt{[CLS]} vector, cosine similarity), since \textmodelname{} has no multi-vector head. We run two checks. Cross retrieval covers all 25 unique retrieval tasks of MTEB (eng, v2) and MTEB (Multilingual, v2); Table~\ref{tab:text-compat-cross} lists seven representative tasks from the full set: four English (ArguAna, SCIDOCS, FiQA2018, CQADupstack Unix), one multilingual (WikipediaRetrievalMultilingual), one mixing code and text (StackOverflowQA) and one Chinese (CovidRetrieval). The agreement check uses these seven tasks.
\begin{itemize}
    \item \textbf{Cross retrieval.} Each task is run four ways: each model on its own index (the two baselines), and both cross combinations, where queries encoded by one model search a corpus encoded by the other. If the cross runs score like \modelname{} on its own index, the vectors can share an index.
    \item \textbf{Same-text agreement.} For up to 2{,}000 queries and 2{,}000 documents sampled per task, we compare the vectors the two models produce for the same text: the mean cosine between a text's two vectors (\emph{same cos}), the mean cosine between different texts across the two models (\emph{other cos}, the floor), and the share of texts whose nearest neighbor in the other model's space is their own twin (\emph{twin@1}). We also report twin@1 after an orthogonal (Procrustes) map fitted on one half of the sample and tested on the other (\emph{rot.\ twin@1}). A high rot.\ twin@1 with a low twin@1 would mean that the spaces have the same shape but are rotated, so a fixed linear map, but not the raw vectors, would make them compatible.
\end{itemize}
Both models run in bf16, with mteb 2.27.14 and sentence-transformers 6.3.0, and scores are the MTEB leaderboard's nDCG@10. For the pooled differences we give 95\% bootstrap confidence intervals over queries.

\begin{table}[h]
    \centering
    \small
    {\renewcommand{\arraystretch}{1.2}
    \begin{adjustbox}{max width=\textwidth}
    \begin{tabular}{ll|cc|cc}
    \toprule
        & & \multicolumn{2}{c|}{Own index (baselines)} & \multicolumn{2}{c}{Cross} \\
        Task & Language & \textmodelname{} $\rightarrow$ \textmodelname{} & \modelname{} $\rightarrow$ \modelname{} & \modelname{} $\rightarrow$ \textmodelname{} & \textmodelname{} $\rightarrow$ \modelname{} \\
        \midrule
        ArguAna & English & 0.572 & 0.544 & 0.542 ($-$0.002) & 0.558 ($+$0.014) \\
        SCIDOCS & English & 0.219 & 0.197 & 0.201 ($+$0.004) & 0.204 ($+$0.007) \\
        FiQA2018 & English & 0.398 & 0.340 & 0.351 ($+$0.011) & 0.344 ($+$0.004) \\
        CQADupstack Unix & English & 0.453 & 0.410 & 0.411 ($+$0.001) & 0.408 ($-$0.002) \\
        WikipediaRetrievalMultilingual & Multilingual & 0.868 & 0.838 & 0.842 ($+$0.004) & 0.827 ($-$0.011) \\
        StackOverflowQA & Code and text & 0.866 & 0.819 & 0.827 ($+$0.008) & 0.821 ($+$0.002) \\
        CovidRetrieval & Chinese & 0.734 & 0.698 & 0.700 ($+$0.002) & 0.682 ($-$0.016) \\
        \midrule
        Mean (25 tasks) & & 0.621 & 0.592 & 0.598 ($+$0.006) & 0.580 ($-$0.012) \\
    \bottomrule
    \end{tabular}
    \end{adjustbox}}
    \caption{Cross retrieval between the single vectors of \textmodelname{} and \modelname{}, nDCG@10. Columns read queries $\rightarrow$ index. Differences in the cross columns are against \modelname{} on its own index, the setup that a mixed index replaces. The task rows are seven of the 25; the last row averages all 25 unique retrieval tasks of MTEB (eng, v2) and MTEB (Multilingual, v2).}
    \label{tab:text-compat-cross}
\end{table}

\begin{table}[h]
    \centering
    \small
    {\renewcommand{\arraystretch}{1.2}
    \begin{tabular}{l|cc|cc}
    \toprule
        Texts & same cos & other cos & twin@1 & rot.\ twin@1 \\
        \midrule
        Queries & 0.922 & 0.691 & 0.997--1.000$^\ast$ & 0.998--1.000$^\ast$ \\
        Documents & 0.932 & 0.712 & 0.999--1.000 & 0.998--1.000 \\
    \bottomrule
    \end{tabular}}
    \caption{Same-text agreement between the vectors of \textmodelname{} and \modelname{}, means over the seven tasks (up to 2{,}000 texts of each kind per task); twin@1 columns give the range over tasks. Rot.\ twin@1 uses an orthogonal map fitted on one half of the sample and is measured on the other. $^\ast$Except ArguAna (0.92, and 0.95 after the map), whose queries include near-duplicate arguments; the same holds between any two of the models we compared.}
    \label{tab:text-compat-agreement}
\end{table}

\paragraph{Results.} Table~\ref{tab:text-compat-cross} shows that queries encoded by \modelname{} retrieve from a \textmodelname{} index at least as well as from their own. Over the 25 tasks the mean rises from 0.592 to 0.598 ($+0.006$, 95\% CI of the difference $[+0.002, +0.009]$), and 19 of the 25 tasks improve or stay level. The one large loss is on TempReasonL1 ($-0.267$), a reasoning task on which \textmodelname{} is a much weaker encoder than \modelname{} (0.113 against 0.391 on their own indexes), so its index is the weak side; even there the cross run stays above \textmodelname{} on its own index. The reverse direction, \textmodelname{} queries against an index built by \modelname{}, loses 0.012 on average ($[-0.016, -0.008]$), but almost all of that comes from TempReasonL1 ($-0.316$); without it, the mean difference is $+0.001$. Compared task by task with the weaker of the two models on its own index, \textmodelname{} queries on a \modelname{} index lose nothing on average ($+0.001$, $[-0.003, +0.005]$). The largest remaining losses are on StatcanDialogueDatasetRetrieval ($-0.050$), WinoGrande ($-0.049$) and the Chinese and multilingual tasks ($-0.016$ on CovidRetrieval, $-0.011$ on WikipediaRetrievalMultilingual). The gap to \textmodelname{} on its own index (0.621 over the 25 tasks) is the multimodal model's lower text quality, not a cost of mixing.

The agreement check (Table~\ref{tab:text-compat-agreement}) shows where the two models differ. A text's two vectors sit at cosine 0.93 for documents but 0.92 for queries: the query path of \modelname{} has moved slightly further from \textmodelname{} than its document path, and the lowest query agreement is on the two tasks where \textmodelname{} queries lose (0.904 on WikipediaRetrievalMultilingual and 0.915 on CovidRetrieval). In every task, however, almost every text's nearest neighbor in the other space is its own twin, and an orthogonal map does not improve this, so the spaces are not rotated copies of each other.

\paragraph{Consequence for deployment.} The direction that works is the one a deployment needs. An existing \textmodelname{} index can stay as it is when page images are added with \modelname{}, and queries encoded by \modelname{} then search the \textmodelname{} text vectors (the \modelname{} $\rightarrow$ \textmodelname{} column) and its own page vectors in one index. Searching an index built by \modelname{} with \textmodelname{} queries also works on most tasks, but loses up to 0.05 on a few, so we recommend encoding queries with \modelname{}.

\paragraph{Limitations.} Only text was compared: whether queries encoded by \textmodelname{} retrieve page images from a \modelname{} index was not tested. The result covers \textmodelname{} only, not granite-embedding-97m-multilingual-r2, and the single-vector head only. The agreement check uses the seven tasks of Table~\ref{tab:text-compat-cross} only. We did not ablate which of the training choices above is responsible for the alignment.

\end{document}